\documentclass[aps,pra,reprint,twocolumn,superscriptaddress]{revtex4}
\pdfoutput=1
\usepackage{color}

\usepackage[pdftex]{graphicx} 
\usepackage{amsmath}
\usepackage{amsfonts}
\usepackage{mathrsfs}
\usepackage{bm}
\usepackage{multirow}
\usepackage{grffile}
\usepackage{braket}
\usepackage{soul}
\usepackage{subfigure}
\usepackage{times}

\begin{document}

\title{Decoherence of an impurity in a one-dimensional fermionic bath with mass imbalance}
\author{A.-M. Visuri}
\affiliation{COMP Centre of Excellence, Department of Applied Physics, Aalto University, FI-00076 Aalto, Finland}

\author{J. J. Kinnunen}
\affiliation{COMP Centre of Excellence, Department of Applied Physics, Aalto University, FI-00076 Aalto, Finland}

\author{J. E. Baarsma}
\affiliation{COMP Centre of Excellence, Department of Applied Physics, Aalto University, FI-00076 Aalto, Finland}

\author{P. T\"{o}rm\"{a}}
\email{paivi.torma@aalto.fi}
\affiliation{COMP Centre of Excellence, Department of Applied Physics, Aalto University, FI-00076 Aalto, Finland}

\begin{abstract}
We study the transport, decoherence and the dissipation of the kinetic energy of a mobile impurity interacting with a bath of free fermions in a one-dimensional lattice. Numerical simulations are made with the time-evolving block decimation method, starting from a state where the impurity and bath are decoupled. We introduce a mass imbalance between the impurity and bath particles and find that the fastest decoherence occurs for a light impurity in a bath of heavy particles. By contrast, the fastest dissipation of energy occurs when the masses are equal. We present a simple model for decoherence in the heavy bath limit, and a linear density response description of the interaction which predicts maximum dissipation for equal masses.
\end{abstract}

\maketitle

\section{Introduction}

Decoherence is the process through which a quantum system transitions to a classical one as it interacts and becomes entangled with an environment. In this process, the quantum coherence between different states of the system, seen as interference patterns in measurements, is destroyed and the superposition state is transformed into a statistical mixture of pointer states \cite{Zurek, Breuer}. Although preparing a system in a quantum coherent state typically requires extreme conditions in order to isolate it from the environment \cite{Wineland, Ketterle, Wieman}, quantum coherent phenomena have also been observed in macroscopic systems at room temperature. Quantum processes can explain the high energy transfer efficiency within photosynthetic complexes \cite{Fleming2010, Jang2013review}; the quantum transport of excitations in biological systems has been observed experimentally with electronic spectroscopy \cite{Engel2007, Panitchayangkoon2010} and photon echo spectroscopy \cite{Scholes2010}. An important question therefore is how the properties of the environment affect the preservation or decoherence of a quantum state. A highly controllable and tunable environment is provided by ultracold atoms trapped with laser beams. For instance, in the experiments \cite{Will, Widera, Ratschbacher}, the system consists of impurity atoms or ions which interact, instead of a thermal bath, with an environment of atoms of a different type. One simple property of the environment that can be varied in such experiments is the (effective) mass of the environment particles.

Current experimental setups allow to control the internal states of the atoms, their density, and the interactions between them. In a recent experiment, an immobile impurity atom was employed as a two-level system to study the decoherence of a qubit interacting with a Bose-Einstein condensate (BEC) \cite{Ratschbacher}. In other experiments, atoms confined to an optical lattice were immersed in a BEC and the Bogolyubov excitations of the BEC mimicked lattice phonons \cite{Grimm, Oberthaler, DeMarco}. The interaction of the impurity with an environment of ultracold atoms can give rise to polaron behavior \cite{Zwierlein, 2DFermiPolarons, Kuhr}, which has been predicted in many theoretical works \cite{Bruderer, Bruun_review2014, Demler, Hofstetter}. Quantum gas microscopes, first implemented for bosonic atoms, have made it possible to track the motion of a single impurity particle in an optical lattice \cite{Kuhr}: The one-dimensional Heisenberg spin model was realized with rubidium atoms, and the propagation of a spin impurity showed both coherent transport and polaron behavior in different interaction regimes of the bath. Recently, quantum gas microscopes for fermions, relevant for this study, have also been implemented \cite{Thywissen2015, Bloch_PauliBlocking2015, Kuhr2015, Greiner2016, Zwierlein2016}. The impurity-bath problem has also been studied theoretically in one dimension \cite{Kamenev, Massel, Oleg2014April, Oleg2014September, Andraschko, Doggen, Visuri, Kantian}, where effects beyond the polaron picture can occur \cite{Zvonarev, Kantian}.

In the framework of open quantum systems, the system of interest is coupled to a large thermal reservoir. The system decoheres and reaches thermal equilibrium with the reservoir due to the coupling. The development of exact numerical techniques for one-dimensional models in particular has lead to the study of non-equilibrium dynamics and thermalization in closed systems \cite{Olshanii, Lauchli, Sirker}, which has also been observed experimentally \cite{Bloch2012}. In this article, we study the dynamics of decoherence and kinetic energy dissipation by simulating the unitary time evolution of a closed system at zero temperature. Here, decoherence refers to the different position (or equivalently momentum) states of the impurity and not e.g. to different internal states. We solve the time evolution of an impurity atom interacting with environment atoms in a one-dimensional lattice system using the numerical time-evolving block decimation method \cite{Vidal, Daley}. The decoherence of the impurity is addressed by calculating its reduced density matrix in position basis. In the following, the words bath and environment are used interchangeably. 

The bath consists of free fermions, and we investigate the effects of a repulsive on-site interaction, filling of the bath, and a mass imbalance between the impurity and bath particles on decoherence. To our knowledge, the dynamical decoherence of impurities is largely unstudied, and especially the effect of mass imbalance on decoherence has not been studied in previous literature. In tilted lattices, the drift of the center of mass of the impurity can be used as a measure of dissipation of energy from the impurity to the bath~\cite{Jaksch}. Here, we use the density changes in the bath and changes in energy to quantify dissipation. The impurity and bath are initially decoupled. We find that, on a time scale defined by the impurity tunneling energy, maximum dissipation occurs when the impurity and bath particles have equal mass, whereas maximum decoherence occurs in the limit of a light impurity and heavy bath particles.

The model and the numerical method are introduced in Sec.~\ref{sec:model}. We present the numerical results in Secs. \ref{sec:transport}-\ref{sec:dissipation}, and give an explanation of some of the characteristics in terms of linear response theory in Sec. \ref{sec:linear_density_response}. Section \ref{sec:transport} discusses the transport of the impurity, which  can be characterized as coherent or incoherent. We show that different mass imbalances lead to different types of transport, and compare the coherent case to a single-particle analytic solution. The decoherence of the impurity is investigated in Sec. \ref{sec:decoherence}, where we calculate the purity of the reduced density matrix of the impurity as a function of time for different mass imbalances. The effect of the interaction strength on decoherence is studied in Sec. \ref{sec:strong_interactions}. For strong interactions, the impurity can form a repulsively bound pair with the bath particles. In Sec. \ref{sec:heavy_bath_limit}, we minimize the effect of doublon formation by a very low filling and study the asymptotic decay rate in the limit of infinitely massive bath particles and strong interactions. A simple model is presented for the decay of purity. The dissipation of energy for different mass imbalances is discussed in Sec. \ref{sec:dissipation}. We find qualitative agreement between the numerical simulations and the integrated dynamic structure factor of the bath, which is discussed in Secs. \ref{sec:dynamic_structure_factor} and \ref{sec:overlap}. Finally, conclusions are presented in Sec. \ref{sec:conclusions}.

\section{Model and the numerical method}
\label{sec:model}

The impurity and bath are described by the Hubbard Hamiltonian 
\begin{align*}
H = H_J + H_U,
\end{align*} 
where the kinetic term with tunneling energy $J$ is 
\begin{align*}
H_J = -\sum_{j \sigma} J_{\sigma} (c_{j \sigma}^{\dagger} c_{j + 1 \sigma} + \text{h.c.})
\end{align*}
and the interaction term with on-site interaction energy $U$ is 
\begin{align*}
H_U = U \sum_j n_{j \uparrow} n_{j \downarrow}.
\end{align*}
We denote the bath fermions by spin up ($\uparrow$) and the impurity by spin down ($\downarrow$), so that $c_{j \uparrow}^{\dagger}$ ($c_{j \uparrow}$) creates (annihilates) a bath fermion and $c_{j \downarrow}^{\dagger}$ ($c_{j \downarrow}$) the impurity, and $n_{j\sigma} = c_{j\sigma}^{\dagger}c_{j\sigma}$ is the number operator. 
We introduce a mass-imbalance between the impurity and the bath via different hopping parameters, $J_\downarrow\neq J_\uparrow$. The tunneling energy is inversely proportional to the mass, so that for example a heavy impurity moving in a light bath corresponds to $J_\downarrow<J_\uparrow$. In an optical lattice, besides the mass of the atom, the tunneling energy depends on the lattice potential. To create an effective mass imbalance, one can have different tunneling energies for the two spin species by adjusting the relative difference in their lattice depths \cite{Griffin2005} or by magnetic field gradient modulation \cite{Esslinger2015}. In the following, we will refer to the mass imbalance in terms of a light or heavy bath.

Initially, $U = 0$, the bath is in the ground state and the impurity is localized at the center of the lattice. At the beginning of the time evolution, the impurity is released and the interaction is changed to $U > 0$. The interaction is fixed to $U = 1J_{\downarrow}$ except for Secs. \ref{sec:strong_interactions} and \ref{sec:heavy_bath_limit} where varying interaction strengths are discussed. The time scale $\frac{1}{J_{\downarrow}}$ is set by the impurity tunneling energy. The numerical time-evolving block decimation (TEBD) method is used for calculating the ground state of the bath and the time evolution of the system. We simulate lattices of size $L = 50$ to $L = 100$, while the Schmidt number used in truncation is $\chi = 100$. A comparison to a higher bond dimension 500 in DMRG simulations shows differences of order $10^{-3}$ in the density matrix elements, which would not be visible in the results shown here. In the real time evolution, we use a time step of $0.01 \frac{1}{J_{\downarrow}}$ or $0.02 \frac{1}{J_{\downarrow}}$.

\section{Transport characteristics for mass imbalance}
\label{sec:transport}

When a particle moves through a medium, the scattering from the surrounding particles often leads to diffusion and a mean-squared displacement which grows linearly in time, $\langle x^2 \rangle \propto t$. More generally, $\langle x^2 \rangle \propto t^{\alpha}$, where $\alpha < 1$ corresponds to subdiffusion and $1 < \alpha < 2$ to superdiffusion. The case $\alpha = 2$ corresponds to the ballistic motion of a freely moving object. The classical concepts of diffusive and ballistic motion can also be applied to quantum particles, in which case ballistic transport, also known as quantum walk \cite{Karski, Preiss2015}, is quantum coherent and gives rise to interference effects. Diffusive transport on the other hand is incoherent. 

\begin{figure}[h!]
\begin{center}
	\includegraphics[width=0.49\linewidth]{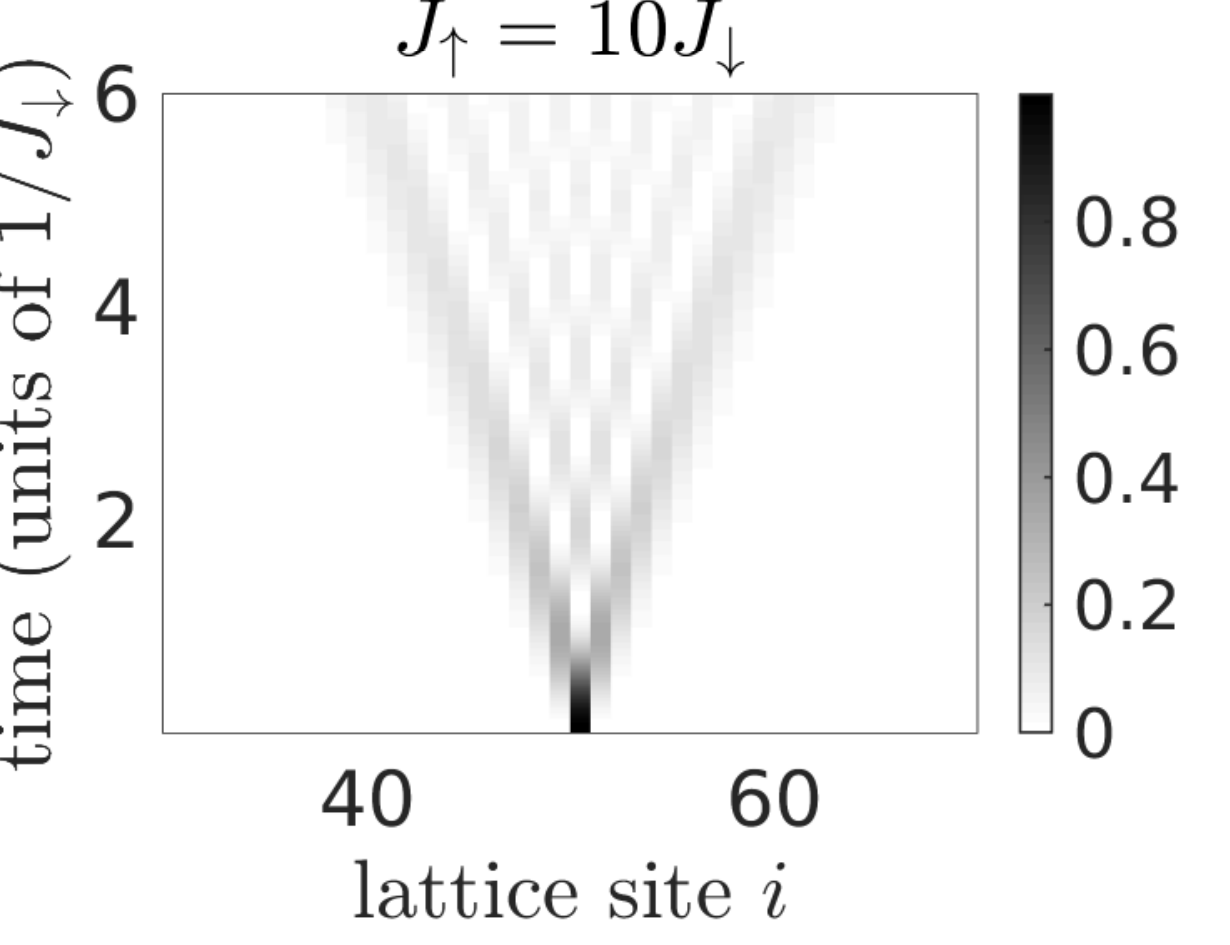}
	\includegraphics[width=0.49\linewidth]{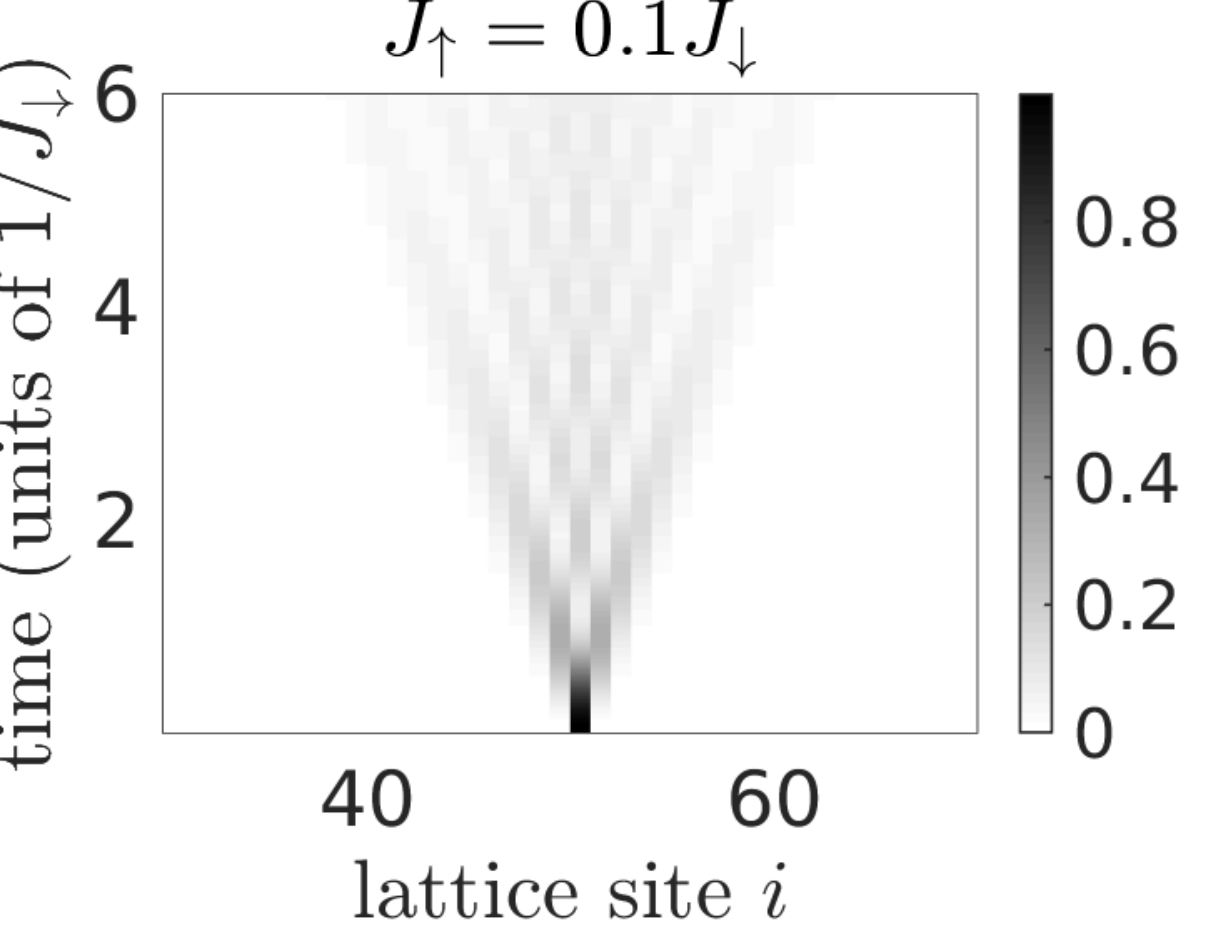}
	
	\includegraphics[width=0.49\linewidth]{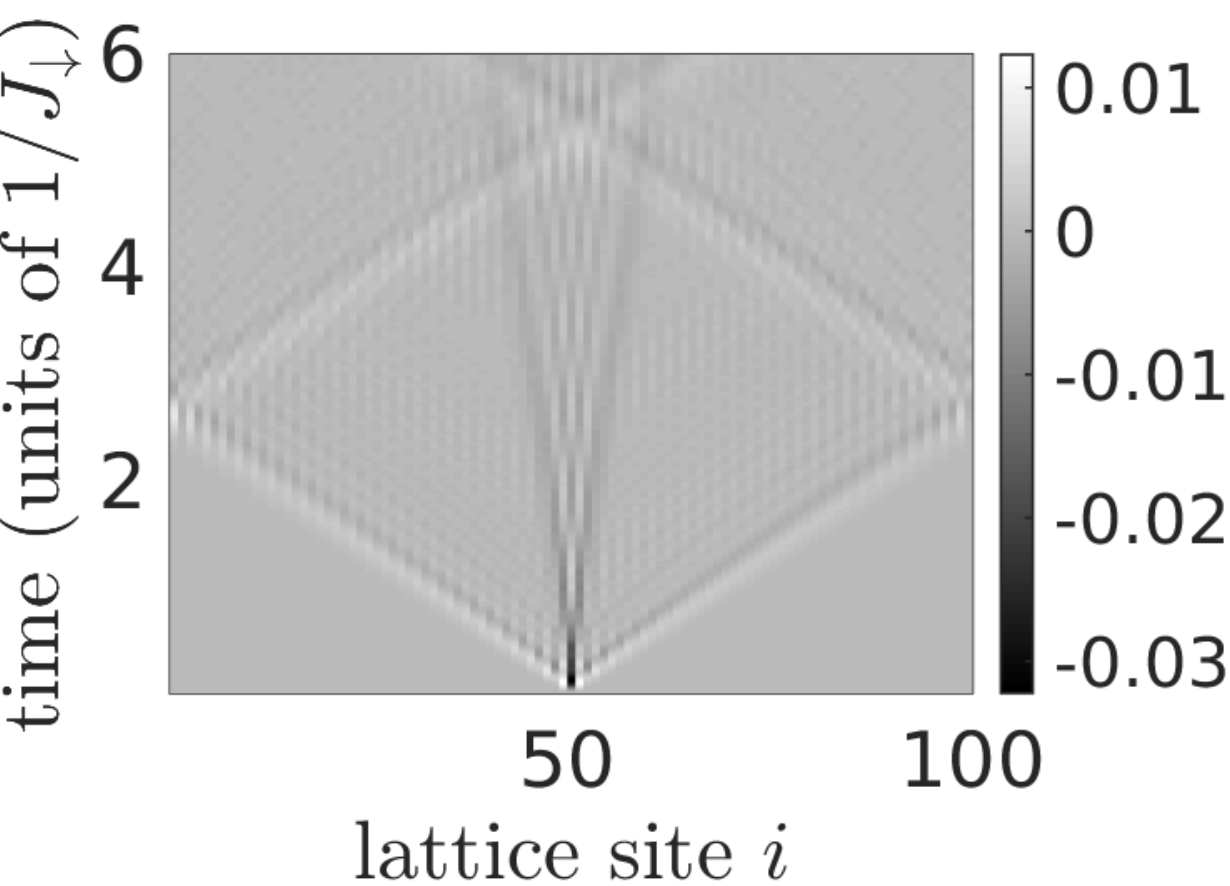}
	\hspace{0.1mm}
	\includegraphics[width=0.49\linewidth]{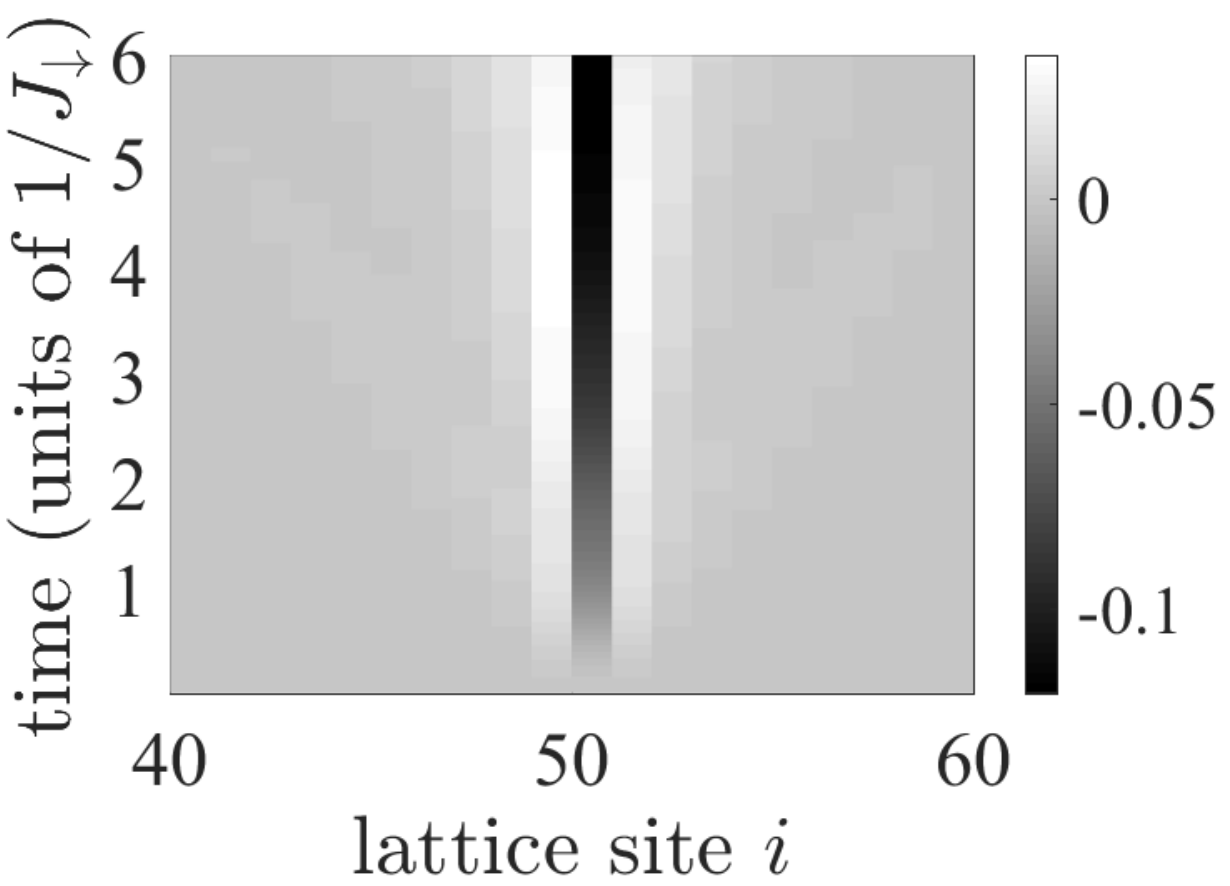}
	\caption{Upper row: The density distribution of the impurity $\langle n_{i \downarrow}(t) \rangle$ as a function of time. For the light bath $J_{\uparrow} = 10 J_{\downarrow}$ (left), the impurity density distribution shows clear wave fronts and an interference pattern, characteristic of coherent transport. For the heavy bath $J_{\uparrow} = 0.1 J_{\downarrow}$ (right), the interference pattern of the impurity is blurred and there is a maximum of density at the center. Lower row: the density difference $\langle n_{i \uparrow}(t) \rangle~-~\langle n_{i \uparrow}(0) \rangle$ of the bath fermions with respect to the ground state. A reversed color scale is used to make the details distinguishable. Density excitations propagate in the bath with the maximum group velocity of noninteracting particles $2 J_{\uparrow}$, as explained in the text. In the right panel, the excitations do not move from the central site within the simulation time due to the very small maximum group velocity $0.2 J_{\downarrow}$. The on-site interaction is $U = 1J_{\downarrow}$ and filling $f = 0.5$.}
\label{fig:impurity_density_imbalance}
\end{center}	
\end{figure}

\begin{figure}[h!]
\begin{center}
	\includegraphics[width=0.6\linewidth]{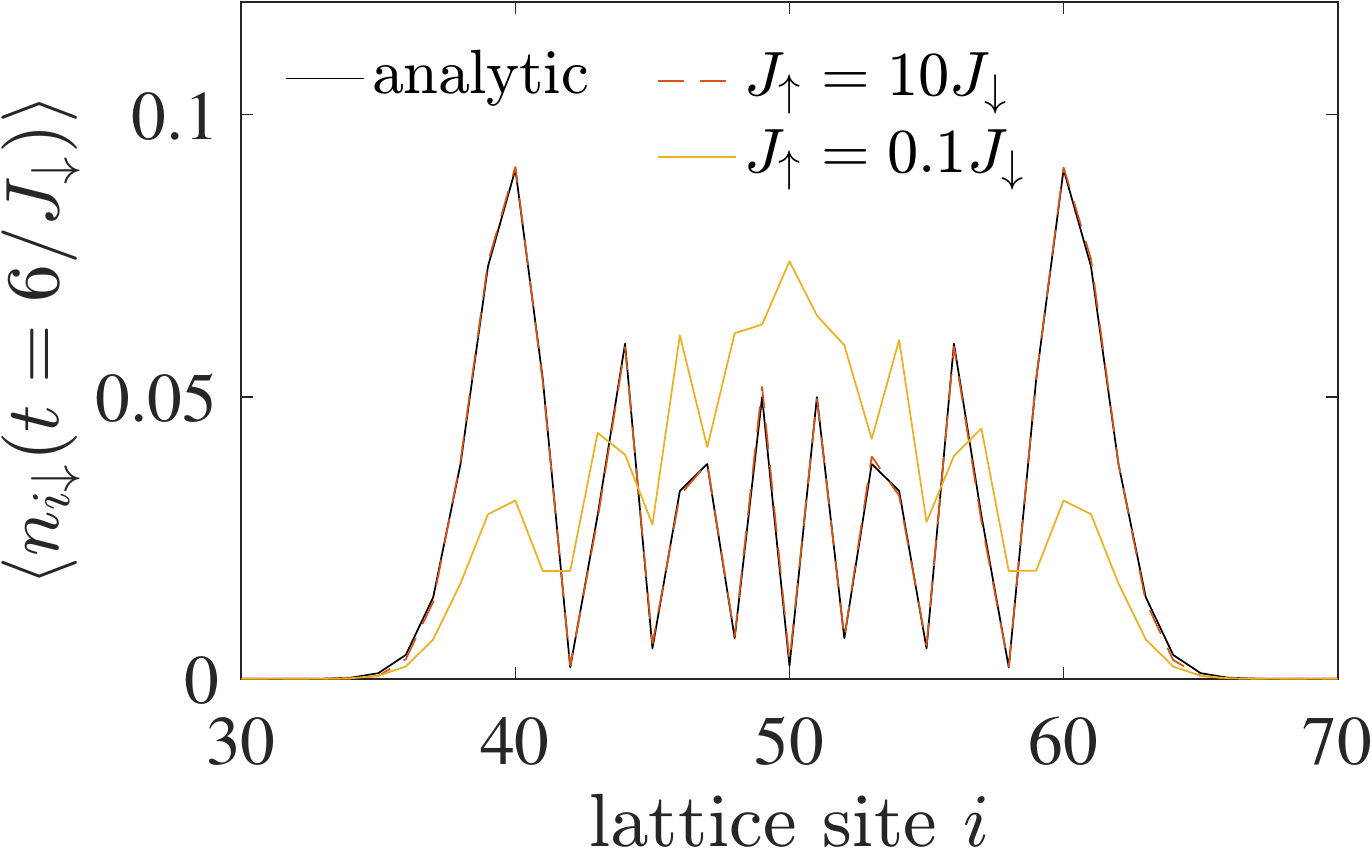}
	\caption{The density profile of the impurity $\langle n_{i \downarrow} \rangle$ at time $t = 6 \frac{1}{J_{\downarrow}}$. The light bath/heavy impurity case ($J_{\uparrow} = 10J_{\downarrow}$) is almost identical to the analytic solution for a free particle with higher-density wave fronts and interference peaks. On the contrary, the density distribution of a light impurity in a heavy bath ($J_{\uparrow} = 0.1J_{\downarrow}$, yellow/light gray line) is peaked at the center.}
\label{fig:line_density}
\end{center}	
\end{figure}

When the propagation of a particle is coherent, interferences produce density minima and maxima, resulting in the density wave fronts and interference patterns seen in Figs. \ref{fig:impurity_density_imbalance} and \ref{fig:line_density}. Similar interference patterns were observed experimentally for a spin impurity propagating in a Mott insulator bath \cite{Kuhr}. 
Figure~\ref{fig:impurity_density_imbalance} shows the densities of the impurity and the bath as functions of position and time for two different mass imbalances. The density profile of the impurity for the same mass imbalances at time $t = 6 \frac{1}{J_{\downarrow}}$ is shown in Fig.~\ref{fig:line_density}. The filling here is $f = 0.5$. For comparison, we show the analytic solution for a single particle, which is almost identical to the density distribution of an impurity interacting with a light bath ($J_{\uparrow} = 10 J_{\downarrow}$). It can therefore be concluded that almost no coherence is lost when the bath is light with respect to the impurity. 

The time evolution of a single particle initially localized in the lattice can be solved analytically by transforming the Hamiltonian $H_{\text{sp}} = -J \sum_{\langle i, j \rangle} c_i^{\dagger} c_j$ into momentum basis. Substituting $c_j^{\dagger} = \sum_k \varphi_{k, j}^* c_k^{\dagger}$, $H_{\text{sp}}$ becomes
\begin{align*}
H_{\text{sp}} = \sum_k \epsilon_k c_k^{\dagger} c_k,
\end{align*}
where $\epsilon_k = -2J \cos(k)$. The above basis functions $\varphi_{k, j} = \sqrt{\frac{2}{L + 1}} \sin(kj)$ are the energy eigenfunctions of a particle in a box, for which $\epsilon_k$ is the same as for the plane wave basis. The initial state is 
\begin{align*}
\ket{\psi(0)} = c_{j_0}^{\dagger} \ket{0} = \sum_k \varphi_{k, j_0}^* \ket{k},
\end{align*}
where $\ket{k} = c_k^{\dagger} \ket{0}$. Time dependence is obtained by operating with the unitary time evolution operator $e^{-i H_{\text{sp}} t}$,
\begin{align*}
\ket{\psi(t)} = \sum_k e^{-i \epsilon_k t} \varphi_{k, j_0}^* \ket{k},
\end{align*}
and for the expectation value of the particle density one obtains
\begin{align*}
\bra{\psi(t)} n_j \ket{\psi(t)} = \left| \frac{2}{L + 1} \sum_{k} \sin(k j_0) \sin(k j) 
e^{i \epsilon_k t} \right|^2.
\end{align*}

\begin{figure}[h!]
\begin{center}
	\includegraphics[width=0.7\linewidth]{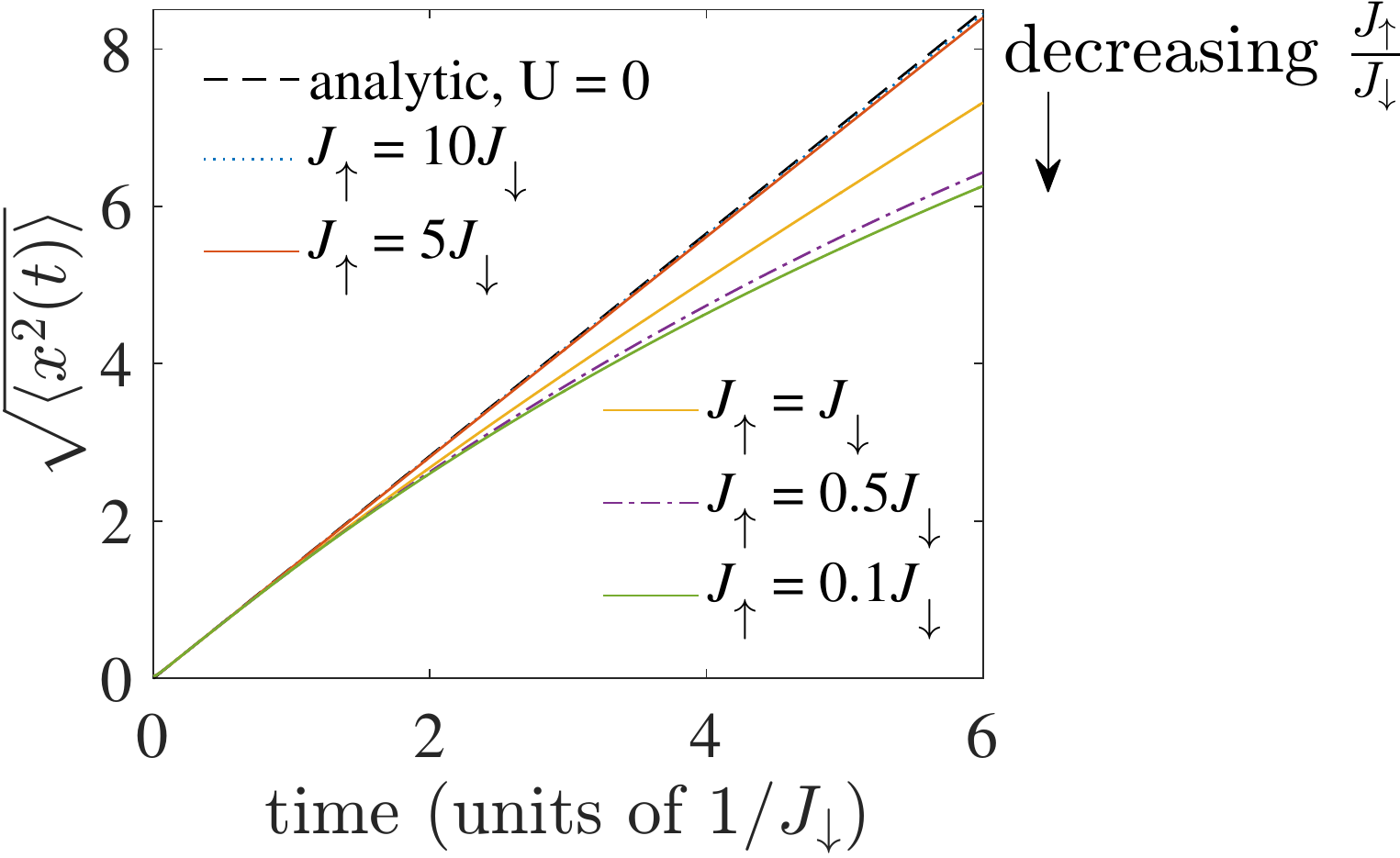}
	\caption{The root mean squared displacement $\sqrt{\langle x^2(t) \rangle}$ as a function of time. The analytic result for a free particle ($U = 0$) grows linearly. The TEBD simulations for varying mass imbalance show that for an impurity interacting with a bath with $U = 1 J_{\downarrow}$, the lines for $J_{\uparrow} = 10 J_{\downarrow}$ and $J_{\uparrow} = 5 J_{\downarrow}$ almost overlap with the analytic result, and the growth decreases from linear for decreasing $J_{\uparrow}/J_{\downarrow}$ (heavier bath).}
\label{fig:msd_aU1}
\end{center}	
\end{figure}

Diffusive propagation is characterized by a maximum of the density distribution at the initial location of the impurity. In Figs.~\ref{fig:impurity_density_imbalance} and \ref{fig:line_density}, one can see that the impurity propagates diffusively in a heavy bath ($J_{\uparrow} = 0.1 J_{\downarrow}$). Density excitations propagate in the bath with the maximum group velocity of noninteracting particles $2 J_{\uparrow}$. In the light bath in Fig.~\ref{fig:impurity_density_imbalance} ($J_{\uparrow} = 10 J_{\downarrow}$), the density excitations propagating at the maximum group velocity can be seen as fast wave fronts, which reflect from the edges of the lattice due to open boundary conditions. Some depletion of density is caused by the impurity, which moves slower. In the heavy bath ($J_{\uparrow} = 0.1 J_{\downarrow}$), the maximum group velocity is very small and the excitations do not move from the central site within the simulation time. 
The change from coherent to incoherent transport with a decreasing mass ratio $J_{\uparrow}/J_{\downarrow}$ is seen in the mean squared displacement $\langle x^2(t) \rangle = \sum_i \langle n_{i \downarrow}(t) \rangle i^2$ in Fig.~\ref{fig:msd_aU1}. The analytically calculated root mean squared displacement for the free particle $\sqrt{\langle x^2(t) \rangle}$ grows linearly in time, whereas in the interacting case with $U = 1 J_{\downarrow}$ the growth slows down for decreasing mass ratio $J_{\uparrow}/J_{\downarrow}$. In order to study the transition into diffusive propagation in more detail, we calculate the reduced density matrix of the impurity and study the purity of the density matrix in Chapter~\ref{sec:decoherence}.

\section{Decoherence of the impurity}

\label{sec:decoherence}

\subsection{The reduced density matrix}

The degree of quantum coherence of a system can be quantified by the purity of its density matrix $\rho$, which is the trace of the density matrix squared, $\text{Tr}(\rho^2)$. The purity has values between 1 and $\frac{1}{N}$, where $N$ is the dimension of the density matrix \cite{Schlosshauer}. For a pure (fully coherent) state, $\text{Tr}(\rho^2) = 1$, whereas an almost classical (mixed) state has a small purity \cite{Weiss}. The impurity-bath system as a whole is in a pure state, and we simulate the unitary time evolution of this total state. Unlike in quantum master equation approaches, the environment here is finite and we simulate it exactly without limitation to weak system-environment coupling or approximations on the memory of the environment. The interaction of the impurity with the bath leads to correlations, due to which the time evolution of the reduced system of the impurity or bath alone is in general not unitary. The reduced density matrix of the impurity is denoted by $\rho_{\downarrow}$ and has elements $\rho_{i, j \downarrow} = \langle c_{i \downarrow}^{\dagger} c_{j \downarrow} \rangle$. Due to entanglement with the bath, $\rho_{\downarrow}$ loses its initial quantum coherence represented by the off-diagonal elements. A quantum gas microscope has recently been applied to directly measure the purity of a many-body state \cite{Greiner2015, Greiner_thermalization}, as proposed earlier \cite{Jaksch2004, Zoller2012, Zoller2013}. A similar technique could allow to monitor the entanglement of an impurity with the environment particles.

\begin{figure}[h!]
\begin{center}
	\includegraphics[width=0.49\linewidth]{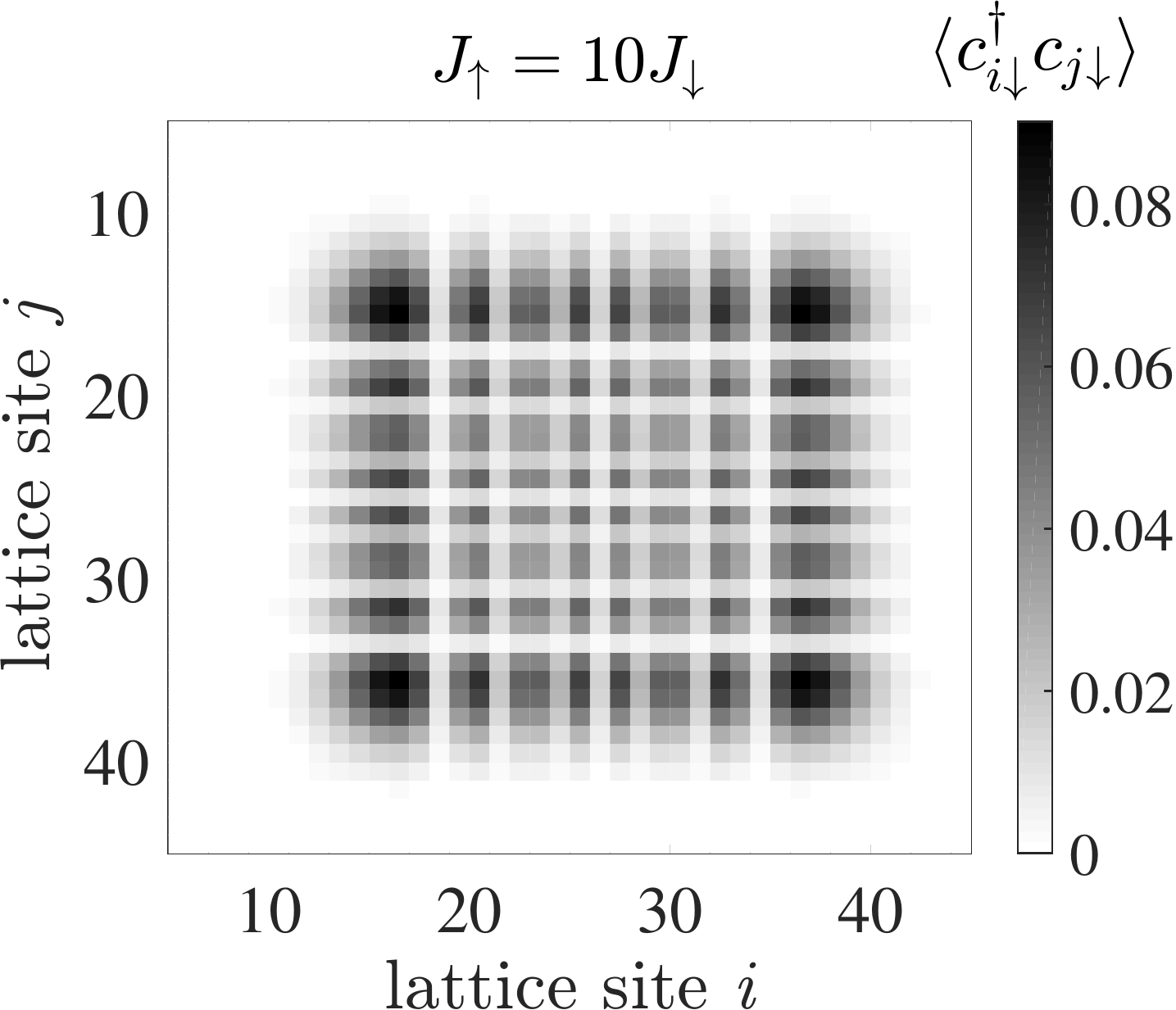}
	\includegraphics[width=0.49\linewidth]{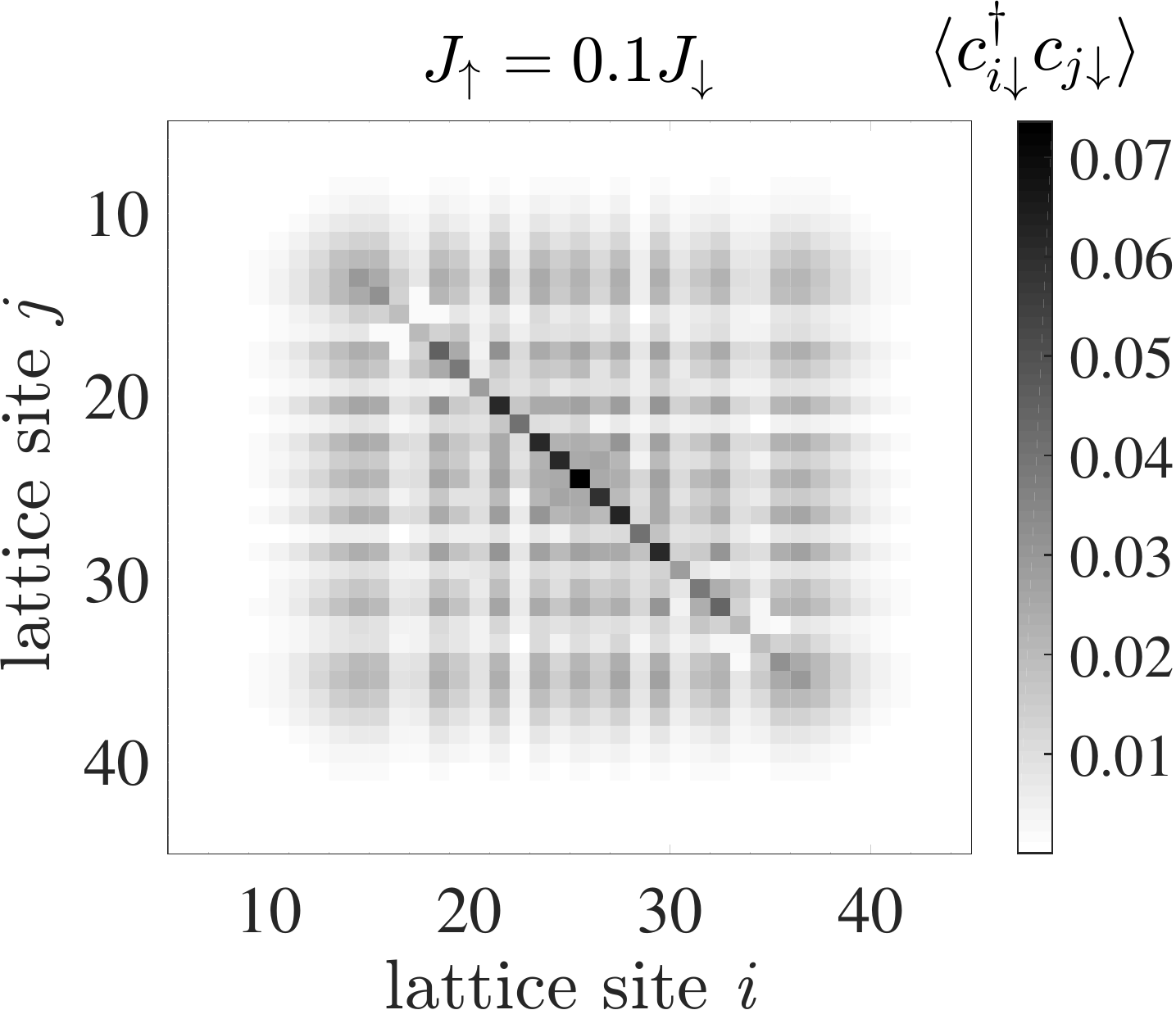}
	\caption{The absolute value of the reduced density matrix of the impurity $|\rho_{\downarrow}|$, where the matrix elements are $\rho_{i, j \downarrow} = \langle c_{i \downarrow}^{\dagger} c_{j \downarrow} \rangle$, at time $t = 6 \frac{1}{J_{\downarrow}}$. In the case of a light bath $J_{\uparrow} = 10 J_{\downarrow}$ (left), the purity of the reduced density matrix is high, $\text{Tr}(\rho_{\downarrow}^2) = 0.99$, whereas for the heavy bath $J_{\uparrow} = 0.1 J_{\downarrow}$, the purity is low, $\text{Tr}(\rho_{\downarrow}^2) = 0.31$.}
\label{fig:impurity_density_matrix}
\end{center}	
\end{figure}

The reduced density matrix of the impurity is shown in Fig.~\ref{fig:impurity_density_matrix} for different values of the mass imbalance at time $t = 6 \frac{1}{J_{\downarrow}}$. In the case of a light bath, the reduced density matrix has large off-diagonal elements and the impurity is in a nearly pure state. The matrix elements differ by approximately $10^{-3}$ from the analytic solution
\begin{align*}
&\bra{\psi(t)} c_i^{\dagger} c_j \ket{\psi(t)} = \\
&\left( \frac{2}{L + 1} \right)^2
\sum_{q, k} e^{i t (\epsilon_q - \epsilon_k)} \sin(q j_0) \sin(q i) \sin(k j_0) \sin(k j)
\end{align*}
for a free particle, which is always in a pure state with $\text{Tr}(\rho^2) = 1$. For an impurity in a heavy bath, the largest values are on the diagonal and the off-diagonal values are small, corresponding to a mixed state. Comparison of Fig.~\ref{fig:impurity_density_imbalance} to Fig.~\ref{fig:impurity_density_matrix} shows that indeed a density distribution with wave fronts and an interference pattern indicates a highly coherent state whereas a distribution peaked at the center is indicative of lower purity. 

\subsection{Purity of the density matrix as a function of time}

\begin{figure}[h!]
\begin{center}
	\includegraphics[width=0.49\linewidth]{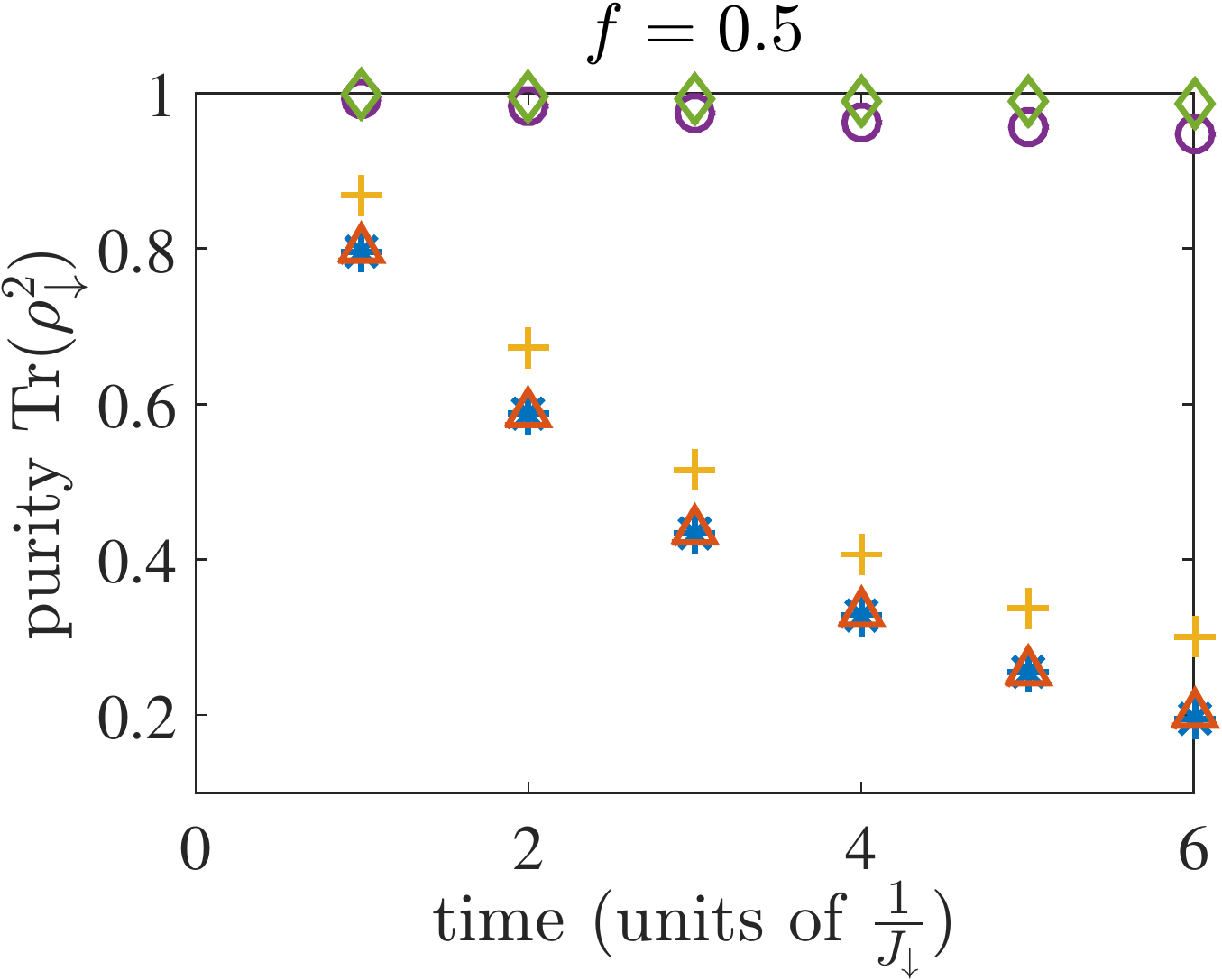}
	\includegraphics[width=0.49\linewidth]{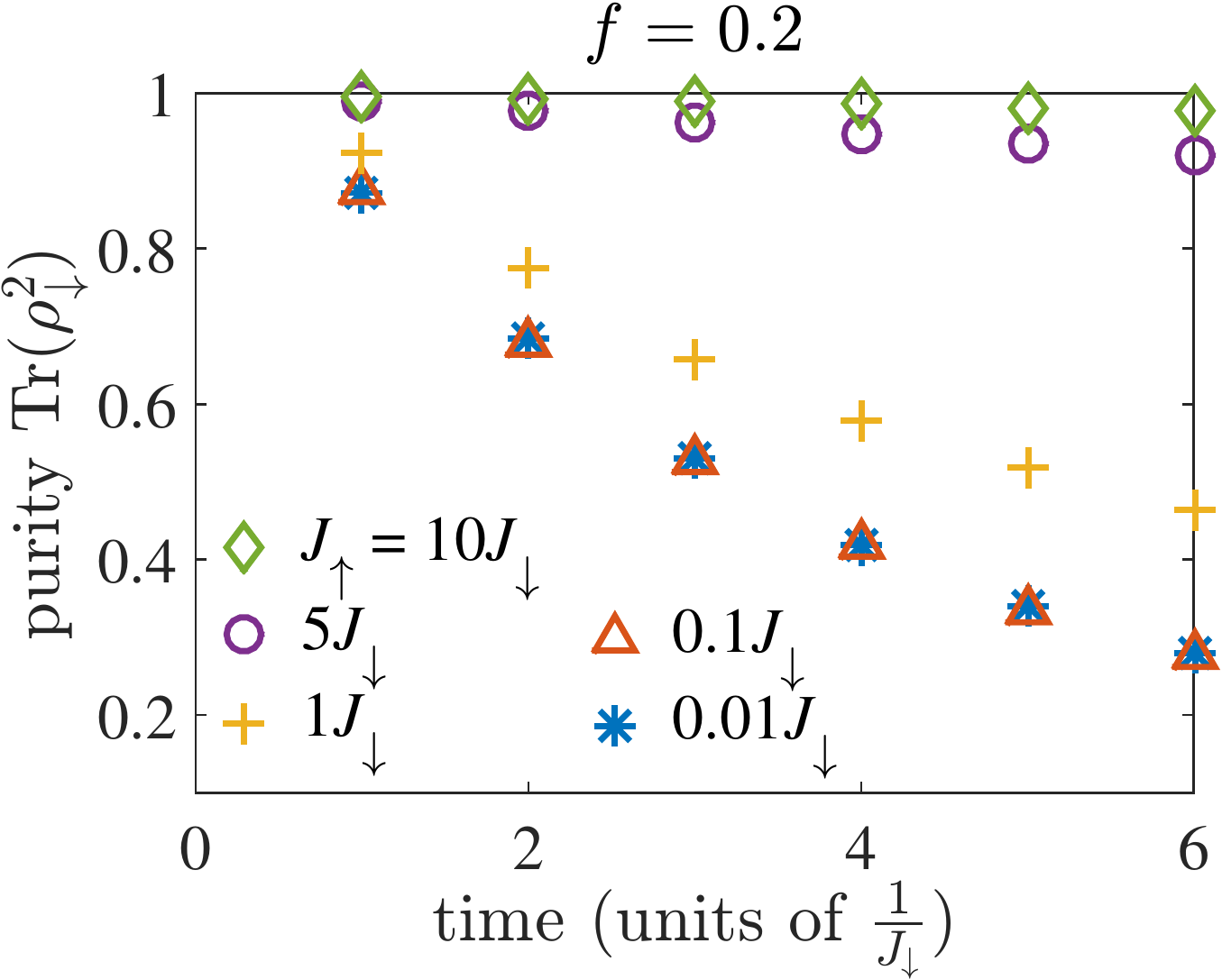}
	\caption{The purity $\text{Tr}(\rho_{\downarrow}^2)$ as a function of time for varying mass ratio with filling $f = 0.5$ of the bath (left) and $f = 0.2$ (right). The decay is faster for heavier baths (smaller $J_{\uparrow}/J_{\downarrow}$). The time unit is $\frac{1}{J_{\downarrow}}$, and the interaction with respect to the impurity hopping is fixed to $U = 1 J_{\downarrow}$.}
\label{fig:purity_time}
\end{center}	
\end{figure}

Figure~\ref{fig:impurity_density_matrix} shows that the mass imbalance between the impurity and the bath has an effect on the decoherence rate: in a heavy bath, the purity decays faster than in a light bath. To study the effect of a mass imbalance on the decoherence more precisely, the purity $\text{Tr}(\rho_{\downarrow}^2)$ is shown as a function of time in Fig.~\ref{fig:purity_time} for varying mass ratios. It can be seen that for an increasingly light bath, the purity stays close to 1 as a function of time. In the opposite case of a heavy bath, the decay rate saturates when $J_{\uparrow} < 0.1J_{\downarrow}$. The limit $J_{\uparrow}/J_{\downarrow} \rightarrow 0$, where the decoherence is fastest, is discussed in Sec. \ref{sec:heavy_bath_limit}. We also vary the filling of the bath and observe that the decoherence is slower as $f$ decreases from 0.5. The purity for filling $f=0.2$, shown in the right panel of Fig.~\ref{fig:purity_time}, is seen to decay slower than for filling $f = 0.5$ (left panel) when $J_{\uparrow} \leq J_{\downarrow}$, which we interpret is due to the impurity interacting with fewer bath particles. A slightly faster decoherence can be seen with the lower filling when $J_{\uparrow} = 5 J_{\downarrow}$ and $J_{\uparrow} = 10 J_{\downarrow}$. We have not found an explanation for this small difference, and understanding it would require further study.

\subsection{Strong interaction}
\label{sec:strong_interactions}

\begin{figure}[h!]
\begin{center}
	\includegraphics[width=0.49\linewidth]{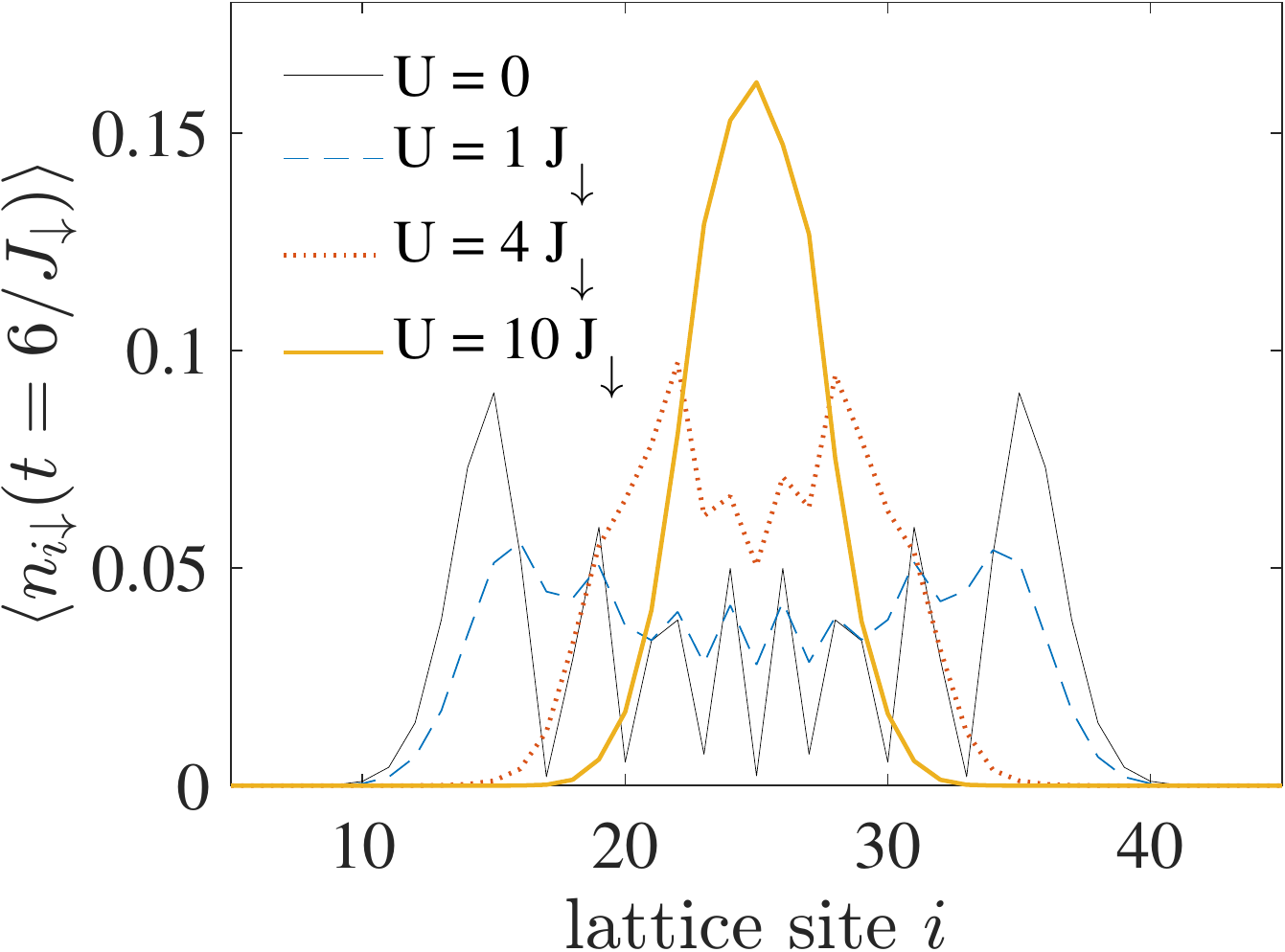}
	\includegraphics[width=0.49\linewidth]{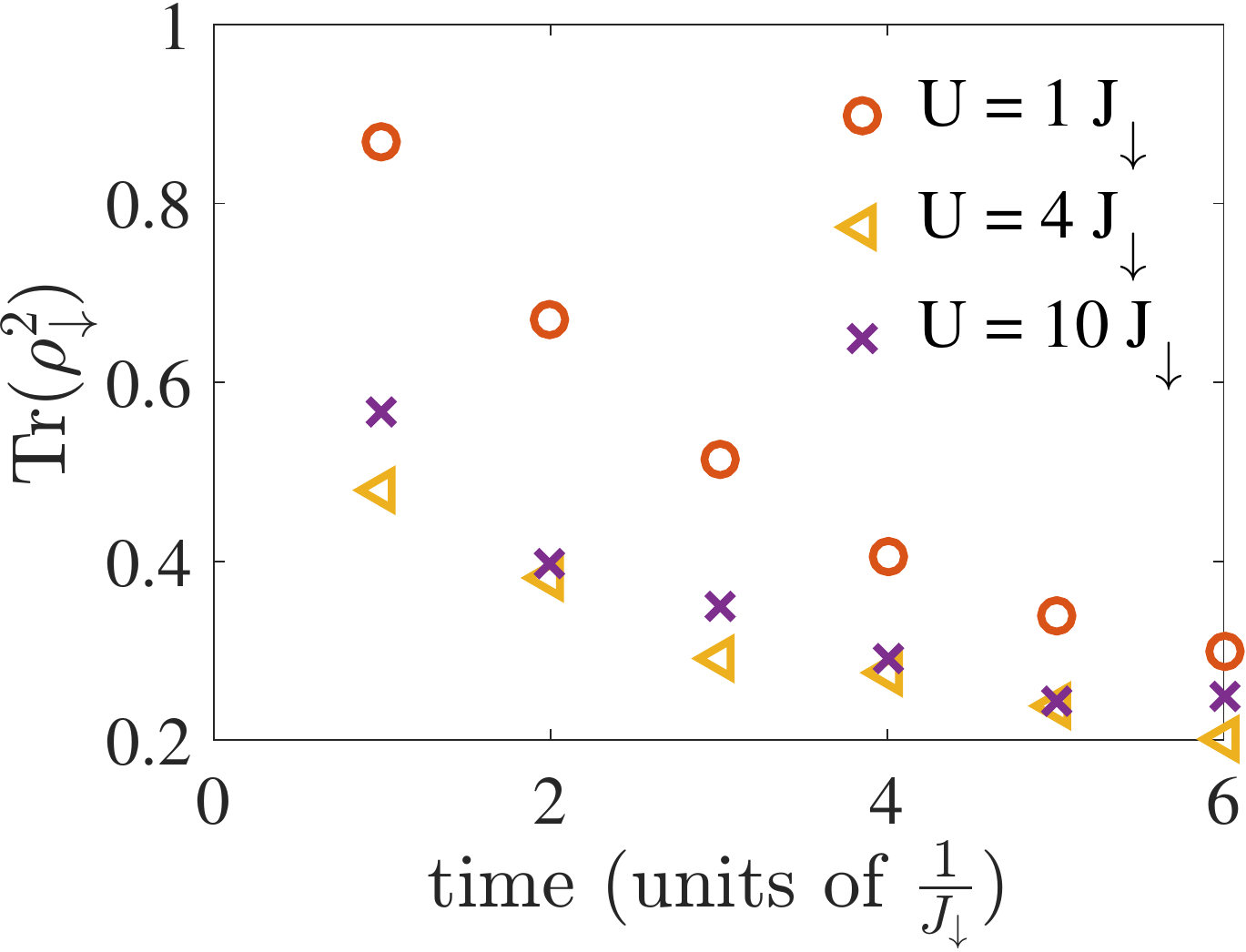}
	\caption{Left: The density profile of the impurity $\langle n_{i \downarrow} \rangle$ at time $t = 6 \frac{1}{J_{\downarrow}}$. The $U = 0$ case is solved analytically as in Fig.~\ref{fig:line_density}, and the $U > 0$ cases numerically with a half-filled bath and equal masses $J_{\uparrow} = J_{\downarrow}$. Right: The purity as a function of time in units of $\frac{1}{J_{\downarrow}}$ for the same parameters ($U > 0$).}
\label{fig:purity_stronger_U}
\end{center}	
\end{figure}

For strong interaction $U \gg J_{\downarrow}$, the impurity can form a repulsively bound pair with a bath particle due to energy conservation.
The maximum group velocity of this doublon is $\frac{4J^2}{U}$, given by the superexchange coupling in a mapping to the Heisenberg Hamiltonian \cite{Georges}. The effect of doublon formation can be seen in Fig.~\ref{fig:purity_stronger_U}. For increasing $U$, the evolution of the impurity wave packet becomes slower and the wave fronts at the edges of the distribution reach a shorter distance within a fixed time. The right panel shows that decoherence is faster for $U = 4 J_{\downarrow}$ and $U = 10 J_{\downarrow}$ than for $U = 1 J_{\downarrow}$.

We interpret that here, two sources of decoherence play a role. The decoherence is partly due to entanglement of the particles forming a doublon, which is larger for stronger interaction, and partly due to scattering, which has a smaller probability for stronger interaction since the bound impurity moves very slowly. A quantum master equation approach has been used earlier to show that the transport of an impurity weakly coupled to a BEC changes from coherent to diffusive with increasing coupling strength \cite{Klein}. The effect of interactions on decoherence was also studied experimentally in a three-dimensional Fermi sea of ultracold ${}^6$Li atoms \cite{Bruun}. In Sec. \ref{sec:heavy_bath_limit}, we investigate the effect of the interaction strength on decoherence when the mass of the bath particles approaches infinity ($J_{\uparrow}/J_{\downarrow} \rightarrow 0$). We use a very low filling, in which case doublon formation can be neglected.

\subsection{The infinitely heavy bath}
\label{sec:heavy_bath_limit}

\begin{figure}[h!]
\begin{center}
	\includegraphics[width=0.5\linewidth]{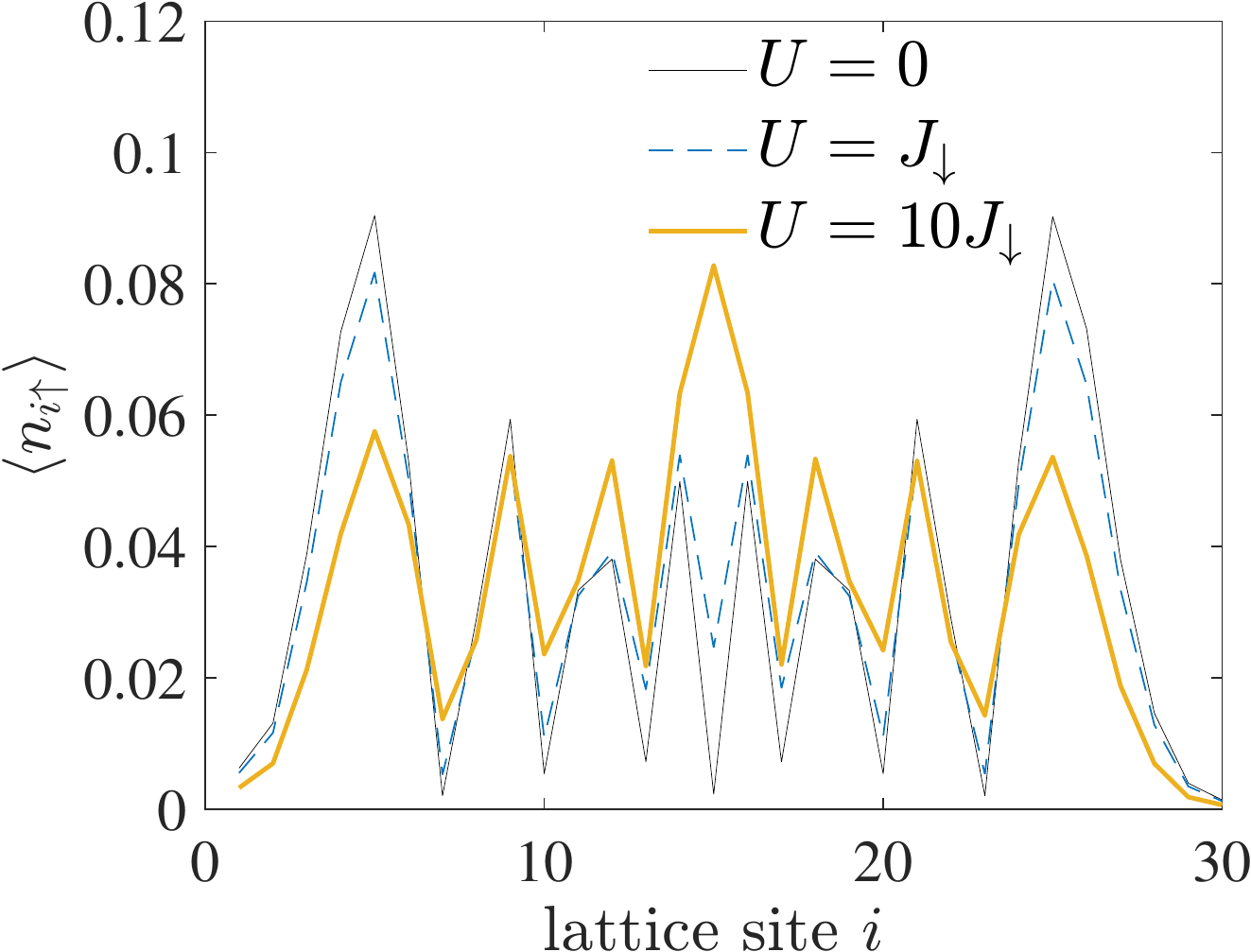}
	\caption{The density profile of the impurity $\langle n_{i \downarrow} \rangle$ at time $t = 6 \frac{1}{J_{\downarrow}}$. The $U = 0$ case is solved analytically and the $U > 0$ cases numerically with filling $f = 0.03$ of the bath and $J_{\uparrow} = 0.01J_{\downarrow}$.}
\label{fig:line_density_2particles}
\end{center}	
\end{figure}

This section focuses on the limit of an infinitely heavy bath $J_{\uparrow}/J_{\downarrow} \rightarrow 0$, where the decoherence of the impurity is fastest, as seen in Fig.~\ref{fig:purity_time}. When $J_{\uparrow}/J_{\downarrow}$ is very small, the environment particles are effectively fixed in place and the impurity experiences a superposition of potentials with fixed barriers of height $U$. Note that a particle in a single realization of a random potential would be a one-body problem with no decoherence, which is quite different from the results we obtain here. At very low filling, the effects of doublon formation can be neglected and the decoherence due to scattering can be analyzed separately. Here we use the lattice size $L = 30$, and the number of spin-up fermions is $N_{\uparrow} = 1$, which gives a filling of the bath $f = 0.03$. We keep here the terminology of an impurity and a bath even though to gain intuition we consider a ''bath'' of only one particle. Figure~\ref{fig:line_density_2particles} shows that the wave fronts at the edges of the impurity density distribution reach the same distance at time $t = 6 \frac{1}{J_{\downarrow}}$ for $U = 10J_{\downarrow}$ as for the noninteracting particle. Unlike in Fig.~\ref{fig:purity_stronger_U}, the slower propagation of the wave fronts characteristic of doublon formation does not occur here. Instead, the density distribution for $U= 10J_{\downarrow}$ is spread out but has a maximum at the center, which is characteristic of diffusive transport due to scattering.

\begin{figure}[h!]  
\subfigure{\includegraphics[width = 0.49\linewidth, trim = 0cm 0.3cm 0cm 0cm]{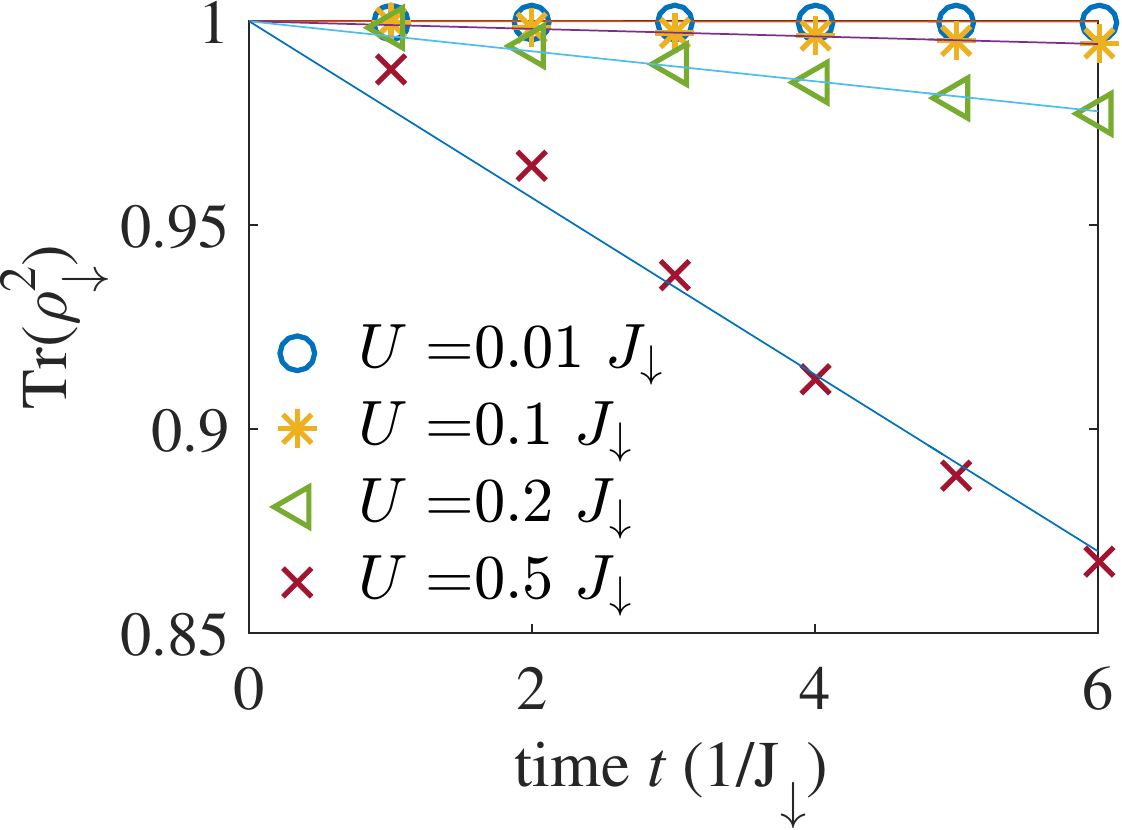}} 
\hspace{0.5ex}  
\subfigure{\includegraphics[width = 0.48\linewidth]{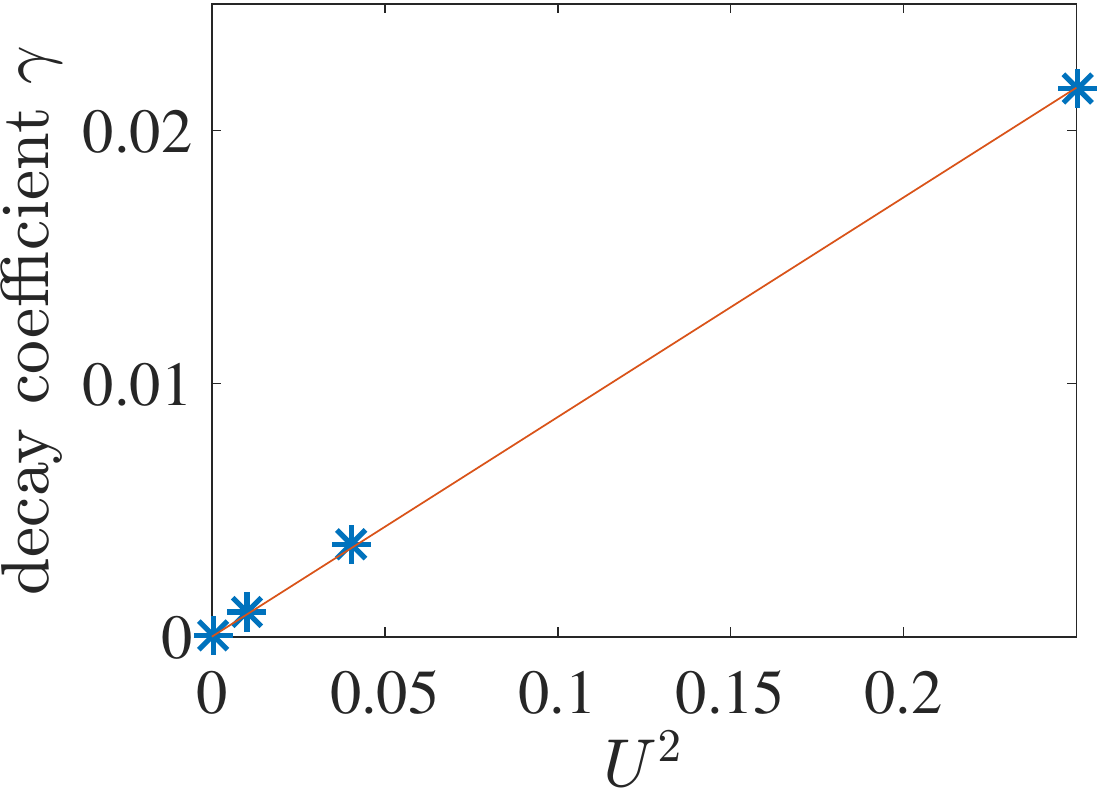}}
\caption{Left: A linear fit $f(t) = 1 - \gamma t$ is made to the purity for $U < J_{\downarrow}$. Right: The decay coefficient $\gamma$ extracted from the fits as a function of $U^2$. A linear fit to these points shows that $\gamma \propto U^2$.}  
\label{fig:decay_coefficient}
\end{figure}

The decoherence rate is related to the transition probabilities in the scattering events. In the perturbative limit of small $U$, one can relate the decoherence rate to the Fermi Golden Rule for transition rates between eigenstates, which are proportional to the square of the transition matrix element. We find that for $U \leq 0.5 J_{\downarrow}$, the decay of $\text{Tr}(\rho_{\downarrow}^2)$ is roughly linear within the time interval studied here, as shown in Fig.~\ref{fig:decay_coefficient}. This would be expected in the short-time limit of an exponential decay. By making a linear fit to the purity, one sees that the decoherence rate $\gamma$ in $f(t) = 1 - \gamma t$ is proportional to $U^2$, as could be expected from the Fermi Golden Rule.

\begin{figure}[h!]
\begin{center}
	\includegraphics[width=0.6\linewidth]{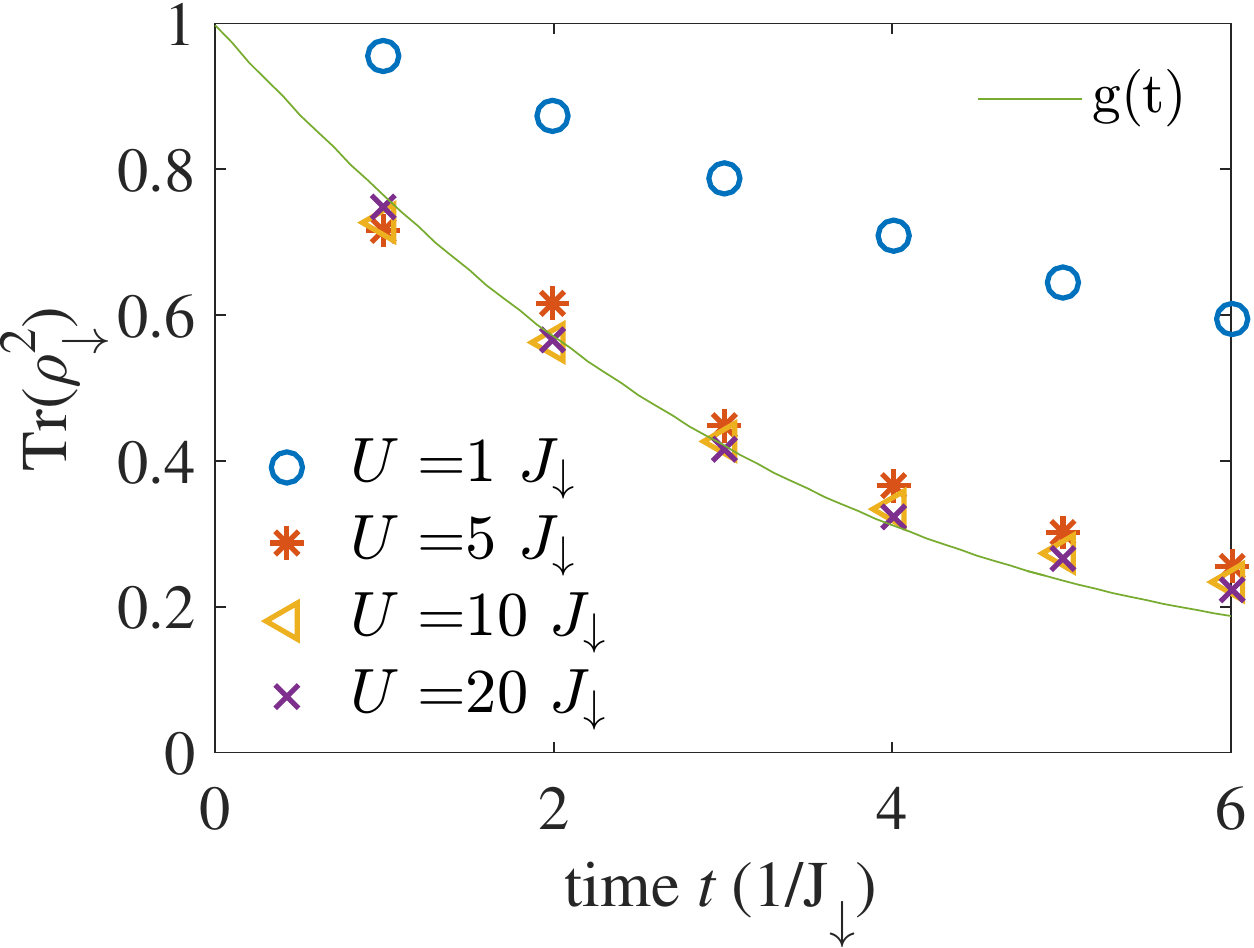}
	\caption{The purity as a function of time for $f = 0.03$ and $J_{\uparrow} = 0.01J_{\downarrow}$. The function $g(t)$ derived in Appendix \ref{app:heavy_bath_limit} agrees well with the simulations.}
\label{fig:two_particles_function}
\end{center}	
\end{figure}

Figure~\ref{fig:two_particles_function} shows the purity for larger $U$. The decoherence is faster for increasing $U$ and the curves saturate for $U \gtrsim 5 J_{\downarrow}$. 
In the large $U$ limit, the decoherence process for two particles can reasonably well be described by a simple model where the impurity propagates as a wave packet with constant velocity and ''measures'' the state of the bath at the lattice site that it reaches. This is a simplification since in reality the time evolution is more complicated, as can be seen from the density profiles above. We find however that this simple model agrees well with the numerical simulations within the time interval shown in Fig. \ref{fig:two_particles_function}. The details of the model are described in Appendix~\ref{app:heavy_bath_limit}. 

\section{Dissipation}
\label{sec:dissipation}

\subsection{Density changes}
\label{sec:density_changes}

\begin{figure}[h!]
\begin{center}
	\includegraphics[width=0.49\linewidth]{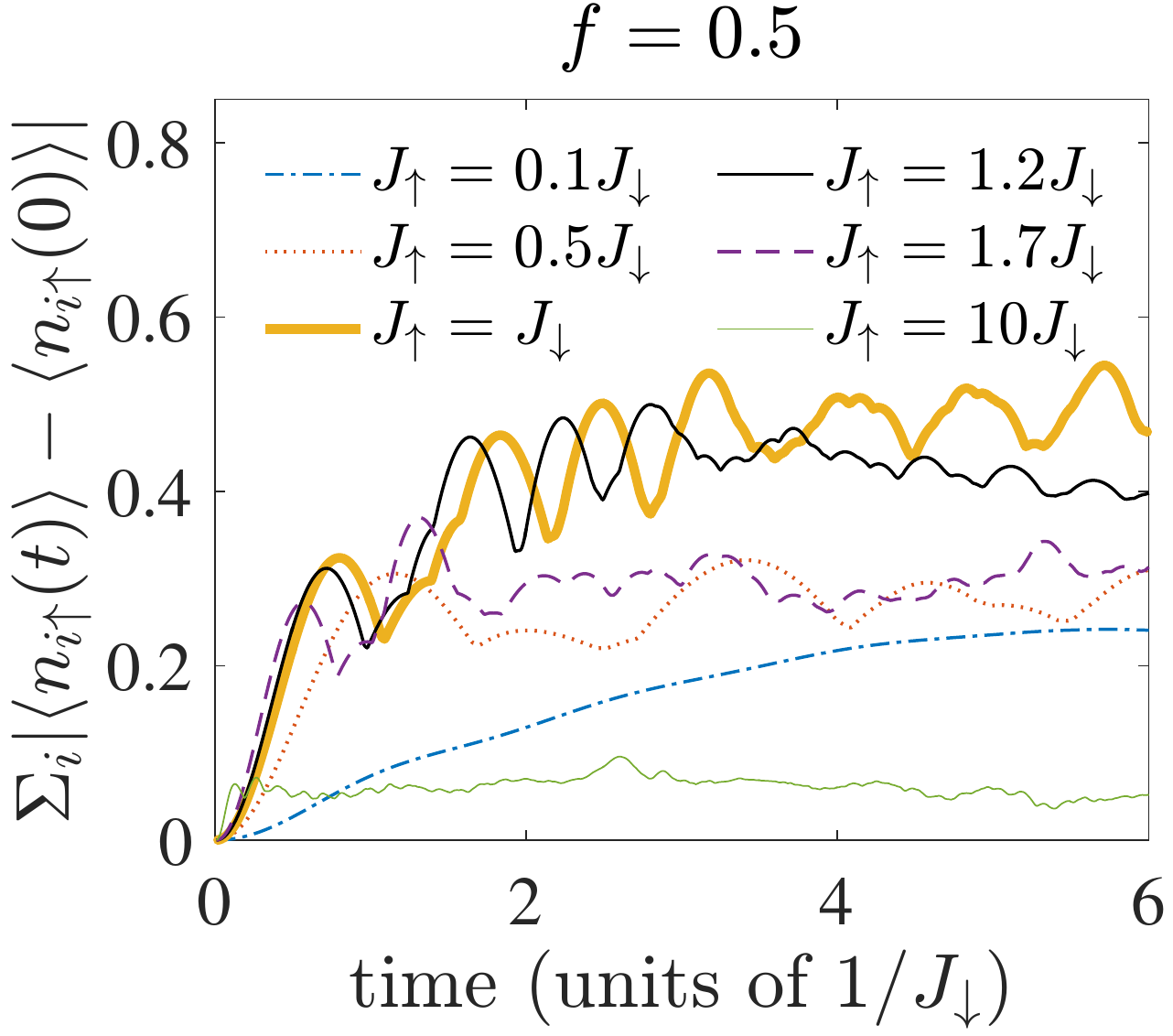}
	\includegraphics[width=0.49\linewidth]{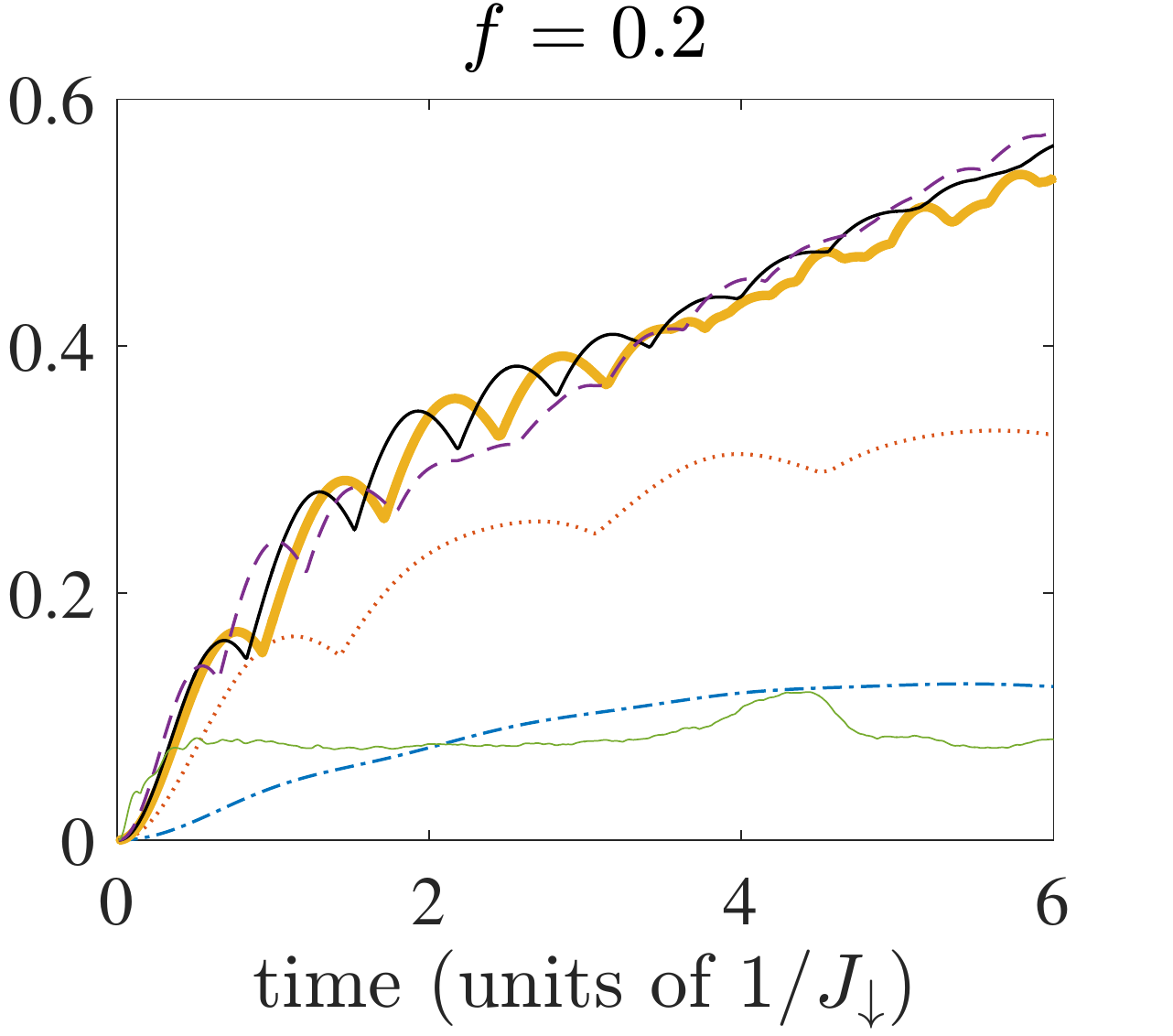}
	\caption{The change in the bath density with respect to the ground state as a function of time, $\sum_{i = 1}^L |\langle n_{i \uparrow}(t) \rangle - \langle n_{i \uparrow}(0) \rangle |$, after the quench $U = 0 \rightarrow 1 J_{\downarrow}$. The mass imbalance varies from $J_{\uparrow}~=~0.1J_{\downarrow}$ (heavy bath, dash-dotted line) to $J_{\uparrow} = 10J_{\downarrow}$ (light bath, thin green/gray line) and the filling is $f = 0.5$ (left) and $f = 0.2$ (right).}
\label{fig:density_change}
\end{center}	
\end{figure}

In addition to decoherence, the interaction of the impurity with the environment leads to the dissipation of energy. In an experiment where impurity atoms moved through a Tonks-Girardeau gas, the dissipation of energy was seen in the widening of the impurity wave packet \cite{Kohl}. In general, decoherence can occur without dissipation, but dissipation or relaxation is always accompanied by decoherence \cite{Schlosshauer}. The dissipation of energy from the impurity to the bath particles can be seen as changes in the bath density profile in Fig.~\ref{fig:impurity_density_imbalance}. Figure \ref{fig:density_change} shows the integrated absolute value of the density change as a function of time, $\sum_{i = 1}^L |\langle n_{i \uparrow}(t) \rangle - \langle n_{i \uparrow}(0) \rangle |$. We find similar results for filling fractions $f = 0.5$, $f = 0.3$ (not shown) and $f = 0.2$. The density change oscillates unevenly and grows at different rates for different tunneling energies of the bath, reflecting the time scale of the displacement of the bath particles and the probability of excitations. The largest density changes occur for masses close to each other $J_{\uparrow} \approx J_{\downarrow}$. The dependence of the probability of creating excitations on mass imbalance is discussed in Sec. \ref{sec:dynamic_structure_factor} and \ref{sec:overlap} in terms of linear response. 

\subsection{Dissipated energy}

\begin{figure}[h!]
\begin{center}
	\includegraphics[width=0.49\linewidth]{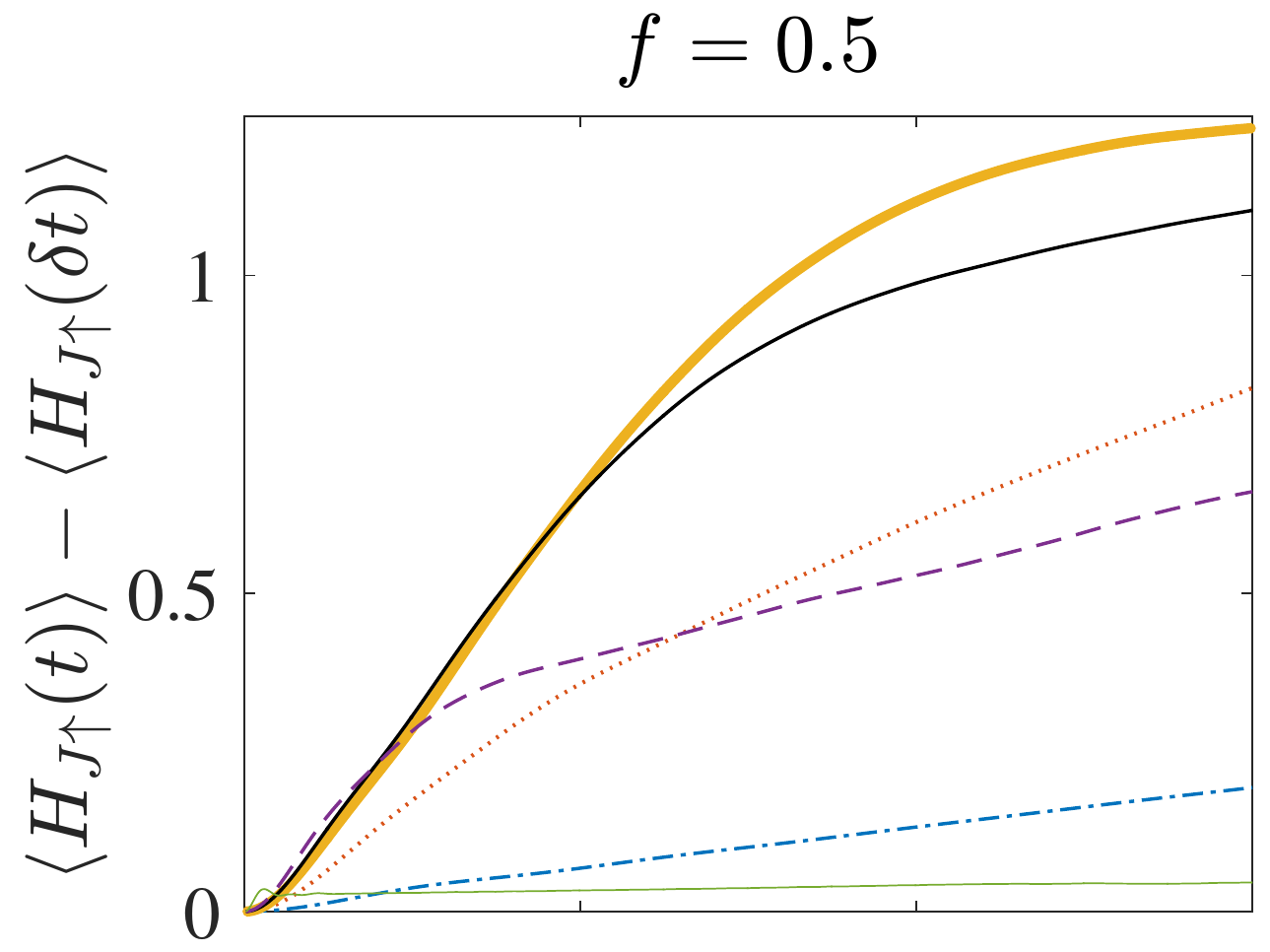}
	\includegraphics[width=0.49\linewidth]{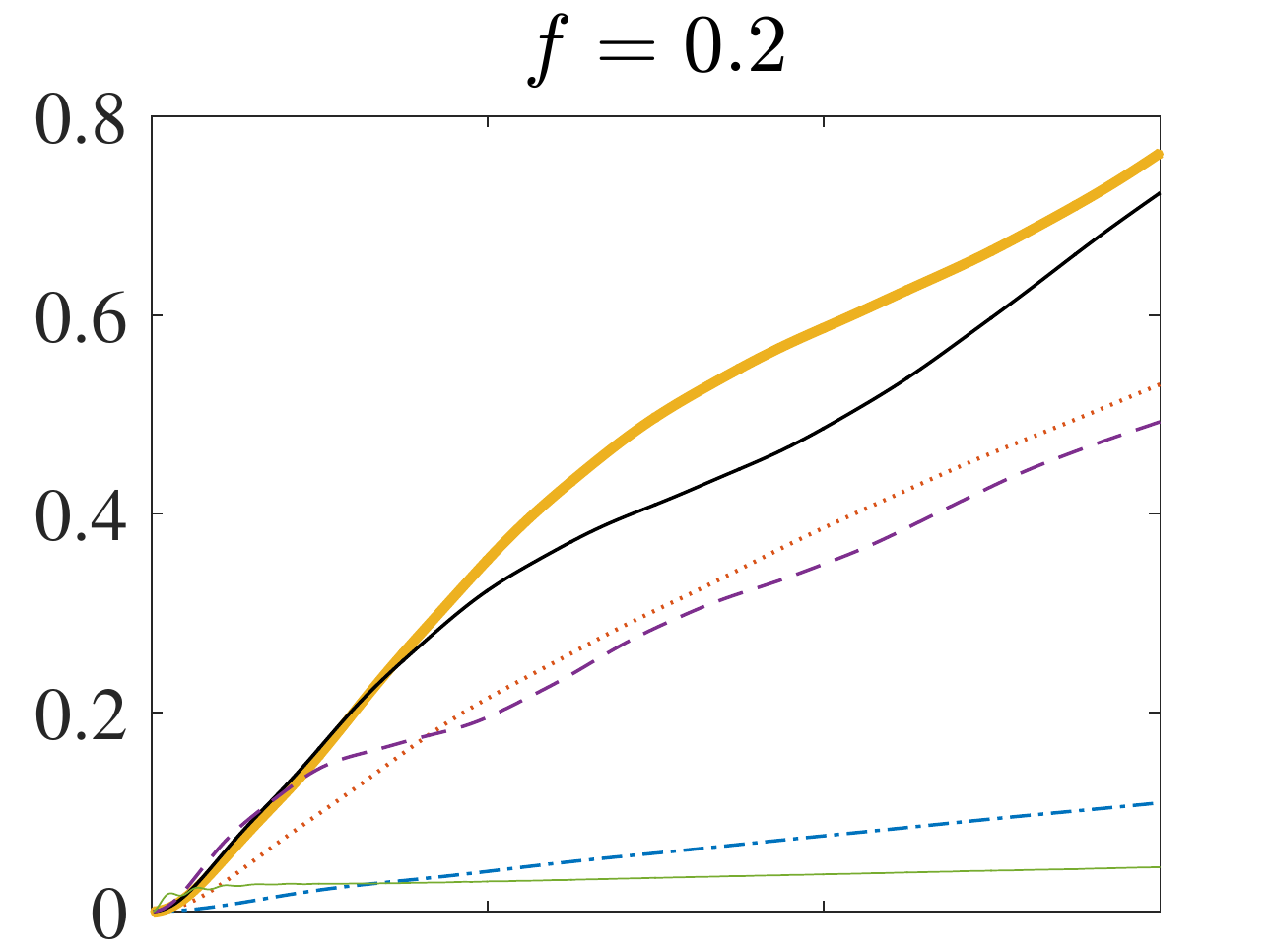}

	\includegraphics[width=0.49\linewidth]{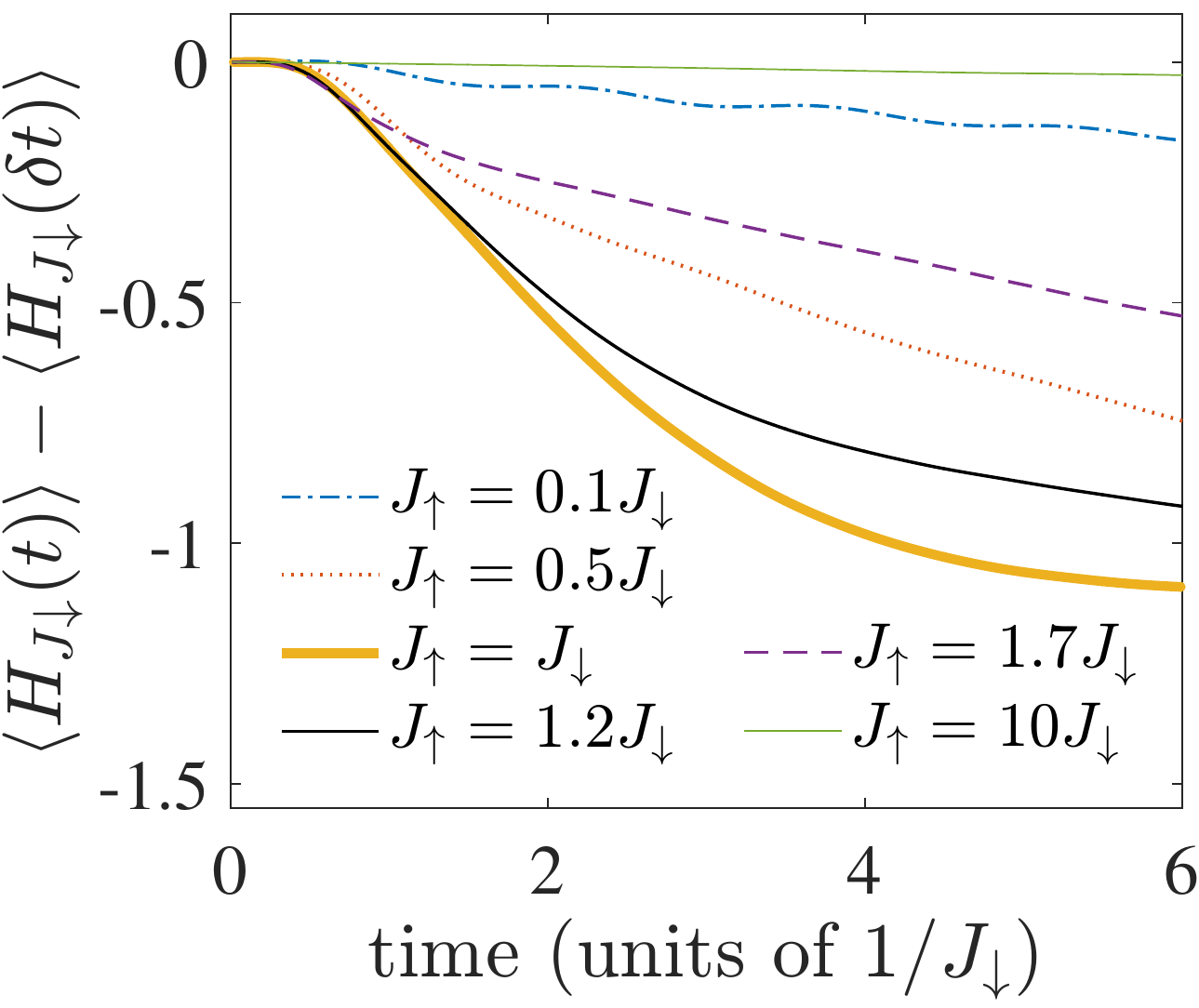}
	\includegraphics[width=0.49\linewidth]{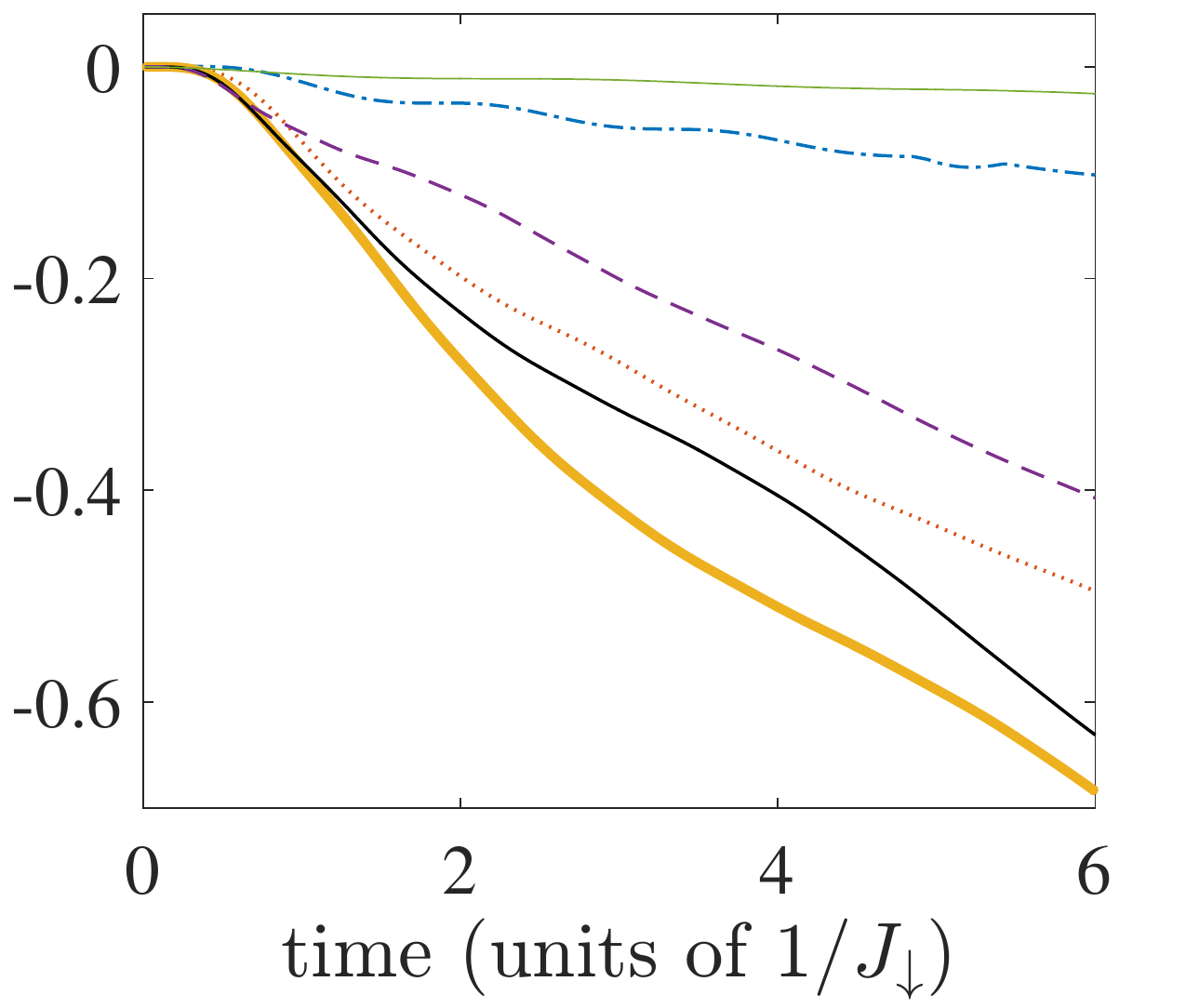}
	\caption{The expectation values of the kinetic terms of the Hamiltonian. For the bath (upper row), $\langle H_{J_{\uparrow}}(t) \rangle$ grows fastest for equal masses, $J_{\uparrow}~=~J_{\downarrow}$ (thick yellow/light gray line), and for the impurity (lower row), $\langle H_{J_{\downarrow}}(t) \rangle$ decreases fastest for equal masses. The initial value at the first time step after the quench $U = 0 \rightarrow 1J_{\downarrow}$ is subtracted in order to compare the changes in energy between different mass imbalances instead of the absolute energies. The filling is $f = 0.5$ on the left and $f = 0.2$ on the right.}
\label{fig:tunneling_energy}
\end{center}	
\end{figure}

\begin{figure}[h!]
\begin{center}
	\includegraphics[width=0.49\linewidth]{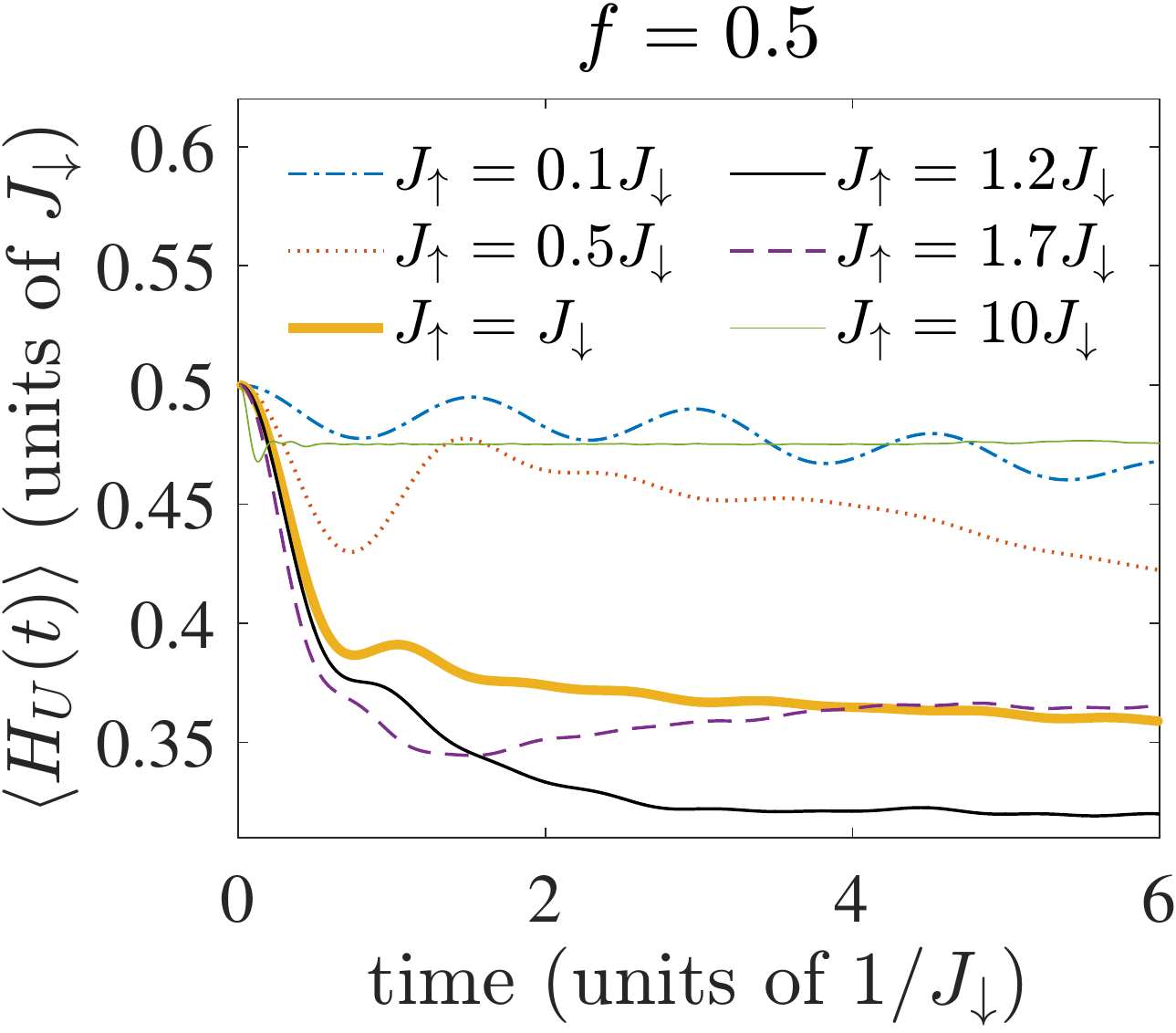}
	\includegraphics[width=0.49\linewidth]{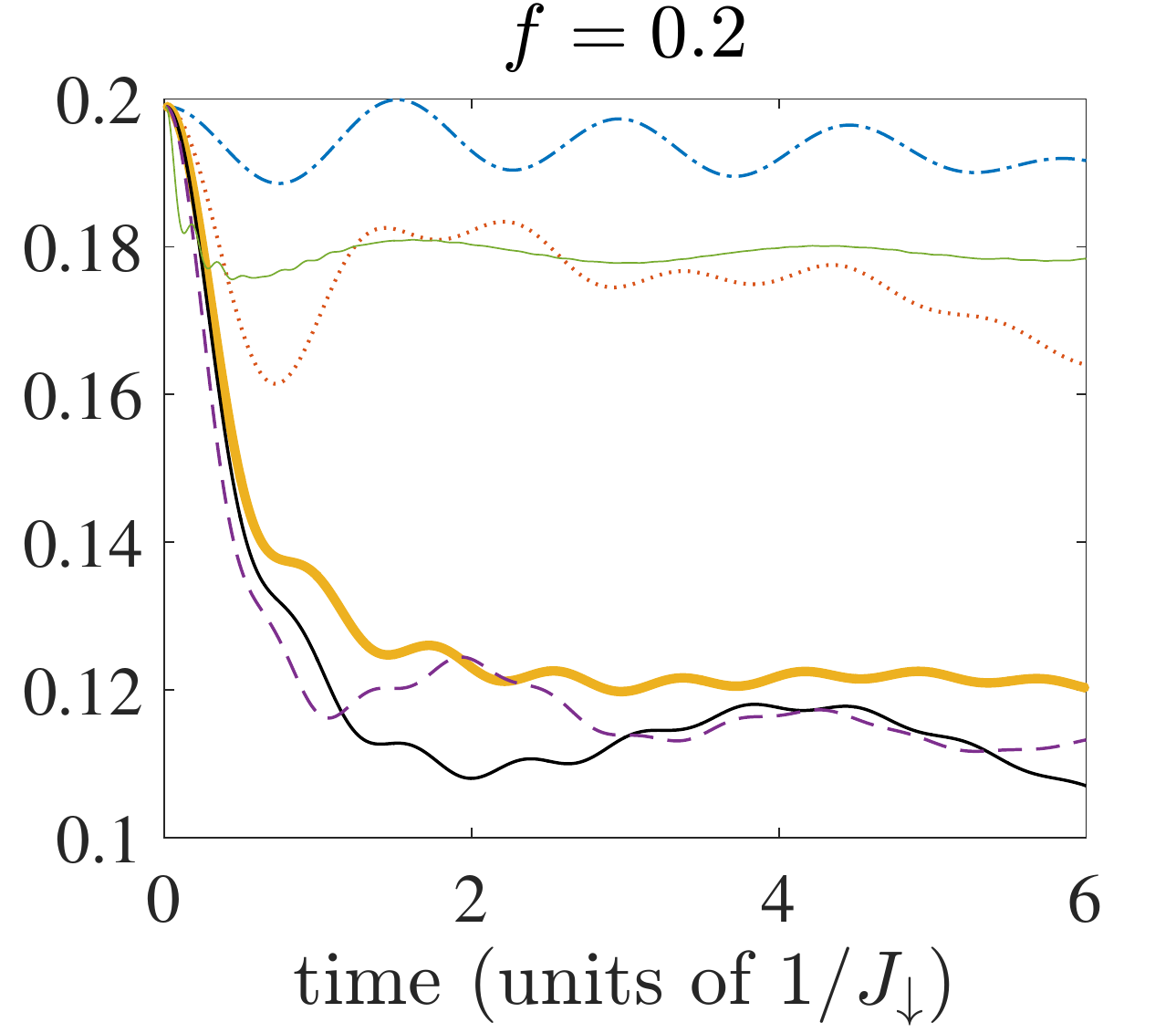}
	\caption{The interaction energy $\langle H_U \rangle$ for $f = 0.5$ (left) and $f = 0.2$ (right.)}
\label{fig:interaction_energy}
\end{center}	
\end{figure}

In order to quantify the dissipation of energy, we show in Fig. \ref{fig:tunneling_energy} the expectation value of the kinetic energy
\begin{align*}
\langle H_{J_{\sigma}}(t) \rangle = -J_{\sigma} \sum_{\langle i, j \rangle} \bra{\psi(t)} c_{i \sigma}^{\dagger} c_{j \sigma} \ket{\psi(t)}
\end{align*} 
as a function of time. The largest increase in the kinetic energy of the bath $\langle H_{J_{\uparrow}}(t) \rangle$ and the largest decrease in the kinetic energy of the impurity $\langle H_{J_{\downarrow}}(t) \rangle$ occur with equal masses, as for the density changes in the bath. The interaction energy shown in Fig. \ref{fig:interaction_energy} is proportional to the total number of doubly occupied sites, 
\begin{align*}
\langle H_U(t) \rangle  = U \sum_i \bra{\psi(t)} n_{i\uparrow} n_{i\downarrow} \ket{\psi(t)}.
\end{align*}
For the heavy $J_{\uparrow} = 0.1 J_{\downarrow}$ and light $J_{\uparrow} = 10 J_{\downarrow}$ bath, the number of doubly occupied sites stays close to the initial value $\langle N_{\uparrow \downarrow}(0) \rangle = f$, as the density distribution of the bath atoms is close to uniform. The largest change in the number of doublons occurs for $J_{\uparrow} \approx J_{\downarrow}$, when there are the largest changes in the bath density distribution. The numerical error in the total energy is shown in Appendix \ref{app:error}. It is of the order $10^{-3}J_{\downarrow}$ within the time interval shown here. It is interesting to note that maximal decoherence of the impurity occurs at the limit of immobile bath particles $J_{\uparrow}/J_{\downarrow} \rightarrow 0$ where the dissipation of energy approaches zero, illustrating that decoherence does not require dissipation.

\section{Linear density response}
\label{sec:linear_density_response}

\subsection{Dynamic structure factor}
\label{sec:dynamic_structure_factor}

\begin{figure}[h!]  
\subfigure{\includegraphics[width = 0.4\linewidth, trim = 0cm -7cm 0cm 0cm]{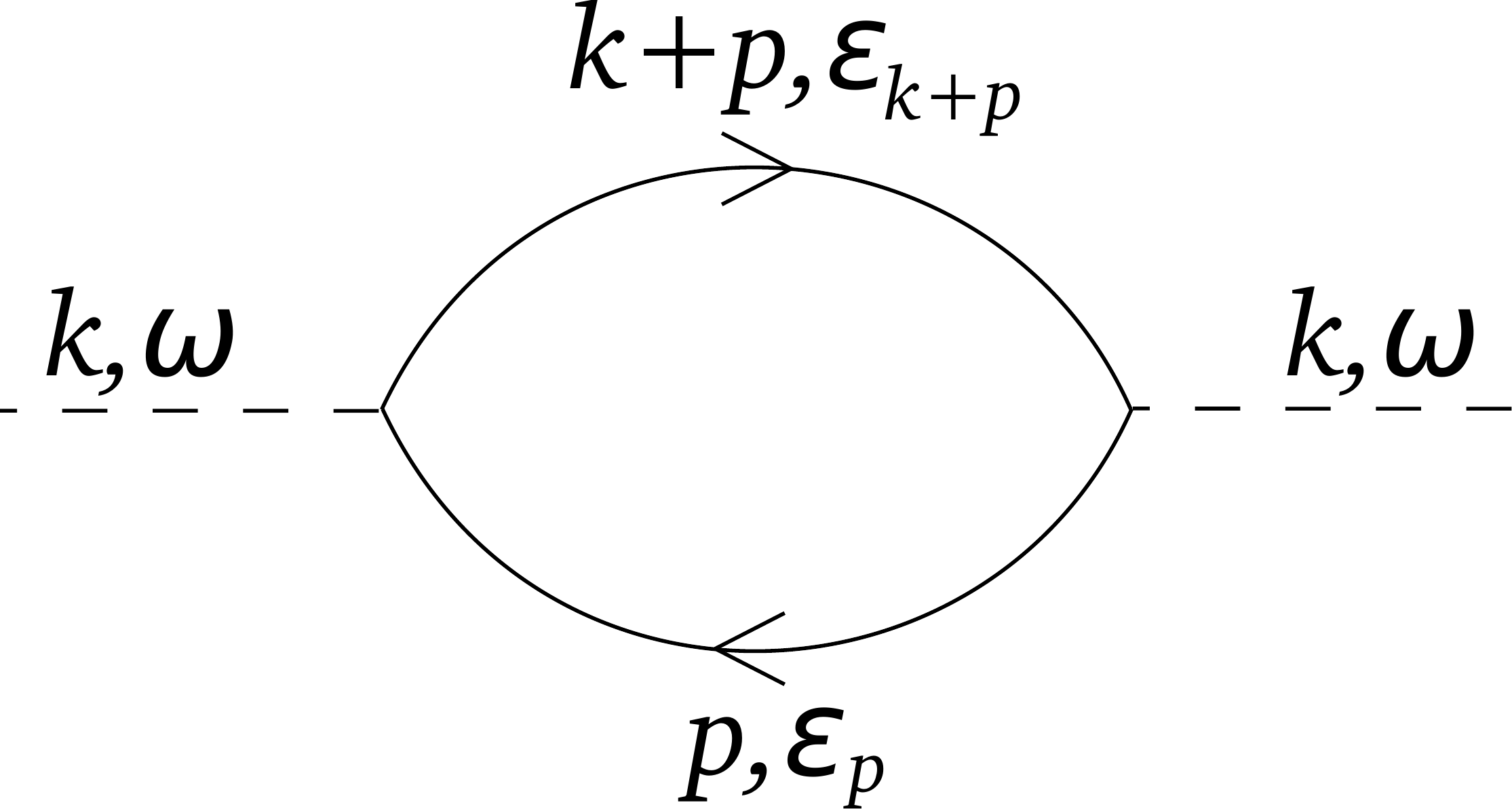}} 
\hspace{2ex}  
\subfigure{\includegraphics[width = 0.5\linewidth]{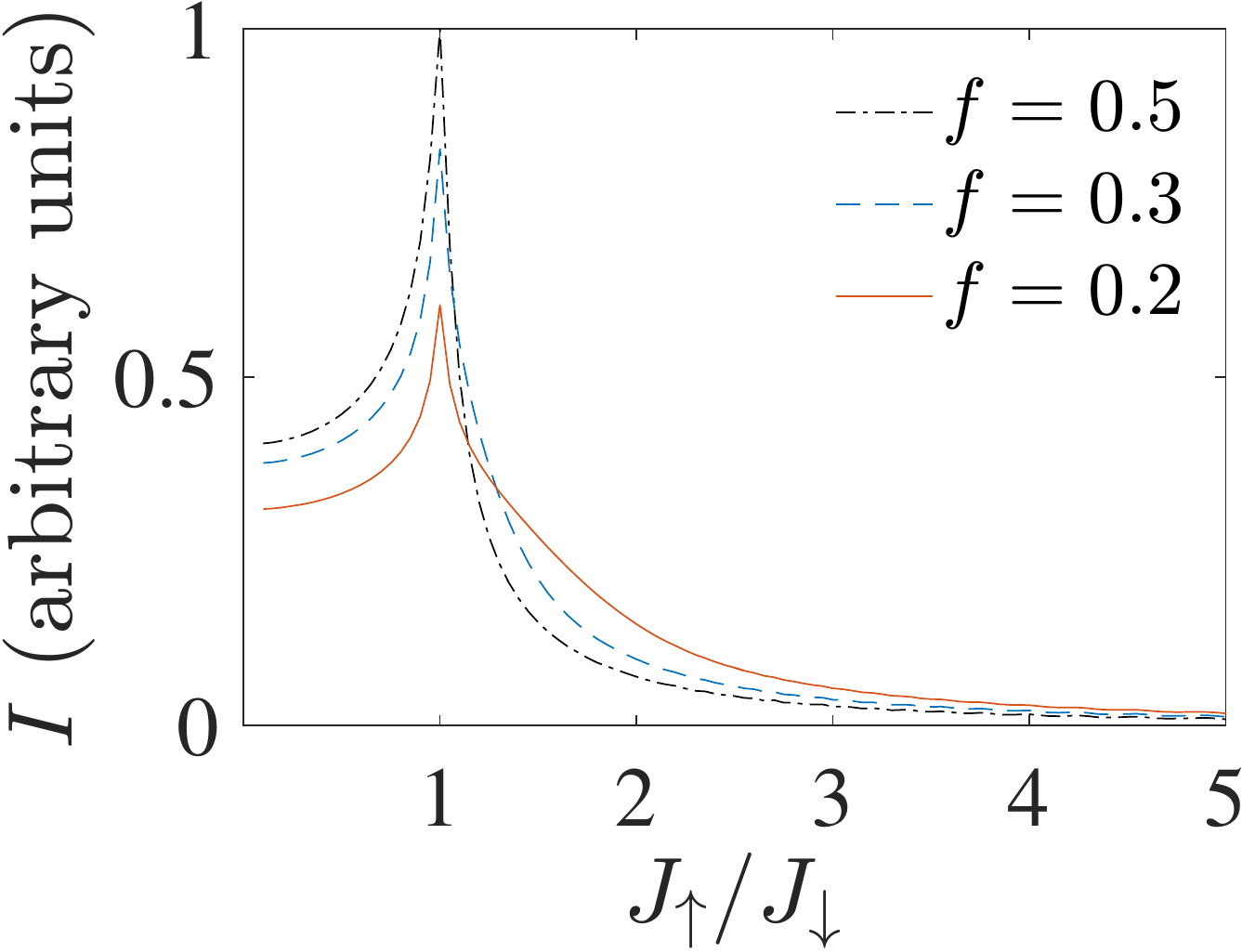}}  
\caption{Left: A particle-hole bubble describing the density fluctuations of the bath. Right: The integrated dynamic structure factor $\mathcal{I}$ as a function of the mass ratio has a peak at equal masses $J_\uparrow = J_\downarrow$, which indicates the largest density response at this value.}  
\label{fig:structurefactor}
\end{figure}

The numerical results on dissipation and density changes in the bath can be understood qualitatively by considering the dynamic structure factor of the bath. The impurity can be viewed as a potential perturbation $V(x, t)$ in the Hamiltonian $H = H_0 + \int V(x, t) n(x) dx$, where $H_0$ is the unperturbed Hamiltonian of the non-interacting bath of fermionic atoms. The linear response in density to the potential $V(x,t)$ is of the form $\langle n(x, t) \rangle = \int \chi(x', x; t', t) V(r', t') dx' dt'$, where the susceptibility $\chi$ is the density-density correlation function in the ground state of $H_0$
\begin{align*}
\chi(x', x; t, t') = -i \theta(t - t') \langle [n(x, t), n(x', t')] \rangle_0.
\end{align*}
In momentum and frequency space, the susceptibility is \cite{Fetter}
\begin{align*}
\begin{split}
\chi(k,\omega) &= \int_{-\pi}^\pi \frac{dp}{2\pi}\, \frac{n_p (1-n_{p+k})}{\hbar \omega + \epsilon_{p}-\epsilon_{p+k} + i\eta} \\
&- \int_{-\pi}^\pi \frac{dp}{2\pi}\, \frac{n_{p+k} (1-n_p)}{\hbar \omega + \epsilon_{p}-\epsilon_{p+k} - i\eta}.
\end{split}
\end{align*}
The first term corresponds to the particle-hole bubble in Fig.~\ref{fig:structurefactor} and the second term is the time-reversed process. At zero temperature, the occupation number is a step function $n_k = \theta(k_F - k)$. The dispersion of the bath particles is $\epsilon_k = -2 J_{\uparrow} \cos(k)$, and $\eta$ is a small imaginary part that acts as a convergence parameter. Physically, $\eta$ describes the lifetime of particle-hole excitations in the bath. Here we use a fixed value $\eta = 0.05\,J_\downarrow$. 

The dynamic structure factor $S(k,\omega) = 2 \mathrm{Im} \chi(k, \omega)$ gives the spectral weight of particle-hole excitations with momentum $k$ and frequency $\omega$. Here, $k$ is the momentum imparted by the impurity atom on the bath and $\hbar \omega$ is the energy transferred in the elastic scattering process. Since the impurity is initially localized in space, and thus occupies all momentum states $q$, the transferred momentum $k$ and energy $\hbar \omega$ are not uniquely defined. Instead, the impurity probes an average quantity, the integrated dynamic structure factor
\begin{align*}
    \mathcal{I} = \int_{-\pi}^\pi \frac{dq}{2\pi} \int_{-\pi}^\pi \frac{dk}{2\pi}\, S\left(k,(\epsilon^\mathrm{imp}_q-\epsilon^\mathrm{imp}_{q-k})/\hbar \right),
\end{align*}
where $q$ is the initial momentum of the impurity, $k$ is the transferred momentum, and $\epsilon^\mathrm{imp}_q-\epsilon^\mathrm{imp}_{q-k}$ is the energy change of the impurity in the elastic scattering process.

Since this simplified description of the impurity-bath interactions only includes single-particle excitations, it does not take into account processes such as doublon formation, which are relevant particularly at short times.
Furthermore, the description of the impurity as a time-dependent perturbing potential is not very accurate, since also the dynamics of the impurity itself is in general changed by the scattering event.
Particularly, the long time limit where the impurity scatters multiple times from different bath atoms cannot be described with a linear response. We find however that the integrated dynamic structure factor, shown in Fig.~\ref{fig:structurefactor}, yields good qualitative agreement with some of the phenomena found in the numerical simulations, as will be discussed below.

\subsection{Overlap of the impurity transitions with the bath excitation spectrum}
\label{sec:overlap}

The largest integrated density changes, observed at equal masses $J_\uparrow = J_\downarrow$ in Fig.~\ref{fig:density_change}, coincide with the peaks in the integrated structure factor. The difference between density changes in the $J_\uparrow = J_\downarrow$ and $J_\uparrow \neq J_\downarrow$ cases is also larger for half-filling than for $f=0.2$, similar to Fig.~\ref{fig:structurefactor}. These results can be interpreted by a simple physical picture where the impurity scatters from bath atoms and creates particle-hole excitations. In particular, in the case $J_\uparrow \approx J_\downarrow$, the particle-hole excitation spectrum of the bath, provided by the dynamic structure factor, overlaps most strongly with the possible transitions of the impurity. This results in the largest density response.

A finite integrated dynamic structure factor $\mathcal{I}$ in the limit of an infinitely massive bath $J_{\uparrow}/J_{\downarrow}~\rightarrow~0$ indicates that a finite number of excitations can be created in this limit. This prediction is explained by the fact that creating excitations does not require energy as the bandwidth of the bath, and thus the range of possible excitation energies, approaches zero. The impurity occupies all momentum states, and it can create all the possible particle-hole excitations in the bath by backscattering. In this process, the impurity is transferred from momentum $q$ to $-q$ and the bath particle gains momentum $2q$. The integrated dynamic structure factor becomes $\mathcal{I} = \int_{-\pi}^\pi dq\, S(2q, 0)$.
These excitations have very little energy, which is in agreement with the decrease in dissipated energy for decreasing $J_{\uparrow}$ seen in Fig.~\ref{fig:tunneling_energy}.

In the opposite limit of a light bath, $J_{\uparrow}/J_{\downarrow} \rightarrow \infty$, the integrated structure factor decays asymptotically as $1/J_\uparrow^2$ and predicts a vanishing response, in agreement with Figs. \ref{fig:density_change} and~\ref{fig:tunneling_energy}. This is due to the reduced overlap of the excitation spectra of the impurity and the bath: The range of energies that the impurity can impart on the bath atoms is limited by the impurity bandwidth $4J_\downarrow$, and the spectral weight of the possible particle-hole excitations in the bath with energy below $4J_\downarrow$ decreases rapidly as the bath hopping $J_\uparrow$ increases. The impurity now has a very small probability of scattering, which is consistent with the coherent transport seen in Figs.~\ref{fig:impurity_density_imbalance}--\ref{fig:msd_aU1} and high purity in Fig.~\ref{fig:purity_time}. 

\section{Conclusions}
\label{sec:conclusions}

We have investigated the transport characteristics, decoherence, and energy dissipation of an impurity propagating in a bath of free fermions. These properties are affected by the mass imbalance between the impurity and bath particles, the strength of impurity-bath interaction, and the filling fraction of the bath. We find, for a fixed interaction strength equal to the impurity tunneling energy, that the transport of the impurity changes from coherent to diffusive as the mass of the bath particles changes from light to heavy with respect to the impurity. To analyze the coherence of the impurity more carefully, we calculate the purity of its reduced density matrix. Similarly to the transport properties, the purity decays faster for increasing mass of the bath particles. A simple model is presented for the case where maximum decoherence occurs, in the limit of infinitely massive bath atoms and strong interactions.

The dynamics studied here could be realized in an experiment with ultracold atoms. The atoms are typically confined by a harmonic trap, and the nonuniform density profile of the bath as well as scattering from the trapping potential could cause effects which are not present in our model. We however expect that the short-time dynamics would show the same essential phenomena. In recent experiments with quantum gas microscopes, sufficiently low temperatures have been achieved to distinguish entanglement entropy, related to the purity, from classical entropy \cite{Greiner2015, Greiner_thermalization}.  Even though the measured purity of the many-body state is less than one, it is possible to distinguish the entanglement between subsystems by comparing their purity to that of the total system. The entanglement entropies of subsystems were measured as a function of time \cite{Greiner2015, Greiner_thermalization}, which gives a promising perspective for the detection of the time-dependent entanglement and decoherence of an impurity particle.

The maximum of dissipation and density changes in the bath is found for equal masses. These results agree with the linear density response, which provides a physical explanation in terms of the overlap of the particle-hole excitation spectrum of the bath with the possible transitions of the impurity. The mass ratio with maximum dissipation could change if there were interactions between the bath atoms. One would expect collective sound mode excitations to become more important in this case, particularly in the case of a superfluid bath. 

Interestingly, we find that maximum decoherence occurs at a different mass ratio than the maximum density response. Dissipation is strongest at the largest overlap of the excitation spectra where the maximal amount of energy can be transferred from the impurity to the bath. Maximum decoherence on the other hand occurs at the limit of infinitely heavy bath atoms, where, due to backscattering, the state of the impurity changes the most even when energy is not transferred. In this case, there is maximal entanglement -- the heavy bath acts like a massive measurement apparatus.

\begin{acknowledgments}
We thank Sebastiano Peotta, Arya Dhar, Juha Kreula, and Oleg Lychkovskiy for useful feedback and our colleagues in the SYNECO project for interesting discussions. This  work  was  supported  by  the  Academy  of  Finland through its Centres of Excellence Programme (2012-2017)  and  under  Project  Nos. 263347, 284621, and 272490, and  by  the  European  Research  Council  (ERC-2013-AdG-340748-CODE). A.-M. V. acknowledges  financial support from the Vilho, Yrj\"o and Kalle V\"ais\"al\"a Foundation. Computing resources were provided by CSC--the Finnish IT Centre for Science and the Aalto Science-IT Project.
\end{acknowledgments}

\appendix

\section{Model for the infinitely massive bath limit}
\label{app:heavy_bath_limit}

An expression for the purity of the reduced density matrix of the bath can be derived by considering a simple model for the decoherence process. 
For a pure state partitioned into two subsystems, the purities of the reduced density matrices of the subsystems are equal, as shown in Appendix \ref{app:Schmidt_decomposition}. Therefore, the same formula gives also the purity of the spin-down fermion. 

When the bath consists of one particle, its ground state can be written as
\begin{align*}
\ket{\psi_{\uparrow}(t = 0)}~=~\sum_j a_j \ket{\uparrow}_j,
\end{align*}
where $a_j = \sqrt{\frac{2}{L}}\sin(\frac{\pi}{L} j)$ and $\ket{\uparrow}_i$ denotes a state where site $i$ is occupied by the spin-up fermion and the other sites are empty. The density matrix $\rho_{\uparrow}(0) = \ket{\psi_{\uparrow}(0)} \bra{\psi_{\uparrow}(0)}$ can be written in matrix form as
\begin{align*}
\rho_{\uparrow}(0) = 
\begin{pmatrix} 
b_{1, 1}	&b_{1, 2}	&\cdots		&b_{1, L}	\\ 
b_{2, 1}	&b_{2, 2}	&			&	\\
\vdots		&			&\ddots		&	\\
b_{L, 1}	&			&			&b_{L, L}
\end{pmatrix},
\end{align*}
where $b_{i, j} = a_i a_j$. The impurity is initially in a superposition of all momentum states, which in the time evolution interfere with each other producing an interference pattern in the density distributions. In a simplified picture, we model the impurity as a delta function wave packet propagating with constant velocity. The group velocity is given by the dispersion relation 
\begin{align*}
v_g(k) = \left| \frac{d\epsilon(k)}{dk} \right| = 2J_{\downarrow} |\sin(k)|,
\end{align*}
and we use here the average velocity
\begin{align*}
\bar{v} = 2J_{\downarrow} \frac{1}{\pi} \int_0^{\pi} \sin(k) dk \approx 1.27 J_{\downarrow}.
\end{align*}
as the velocity of the impurity wave packet. 

The initial state of the impurity is $\ket{\psi_{\downarrow}(0)} = \ket{\downarrow}_{j_0}$, and for a noninteracting impurity, the time-dependent state would in our model be a superposition of wave packets moving to the left and to the right, 
\begin{align*}
\ket{\psi_{\downarrow}(t)} = \frac{1}{\sqrt{2}}\left( \ket{\downarrow}_{j_0 - \bar{v}t} + \ket{\downarrow}_{j_0 + \bar{v}t} \right).
\end{align*}
When the impurity interacts with the bath particle, the total time-dependent state becomes
\begin{align*}
\ket{\psi(t)} = \frac{1}{\sqrt{2}} \left( \ket{\psi_{\text{L}}(t)} + \ket{\psi_{\text{R}}(t)} \right),
\end{align*}
where $\ket{\psi_{\text{L}}(t)}$ is the state in the case where the impurity propagates to the left from $j_0$ and $\ket{\psi_{\text{R}}(t)}$ the state with the impurity propagating to the right. The reduced density matrix of the bath is now
\begin{align}
\begin{split}
\rho_{\uparrow}(t) &= \text{Tr}_{\downarrow}[\rho(t)] = \frac{1}{2}[\rho_{\uparrow \text{L}}(t) + \rho_{\uparrow \text{R}}(t)],
\end{split}
\label{eq:density_matrix_up}
\end{align}
where $\rho_{\uparrow \text{L}}(t)$ is the reduced density matrix of the bath particle in the case where the impurity propagates left and $\rho_{\uparrow \text{R}}(t)$ the reduced density matrix with the impurity propagating right. In the limit of a strong on-site interaction, we assume that the propagating impurity ''measures'' the state of the lattice site that it reaches (or one can equally well think that the bath measures the impurity) so that the state at these sites becomes completely entangled. The state is assumed to be unmodified at the sites that the impurity has not reached.

In matrix form, $\rho_{\uparrow \text{L}}$ is written as
\begin{widetext}
\begin{align*}
\rho_{\uparrow \text{L}}(t) = 
\begin{pmatrix} 
b_{1, 1}				&\cdots	&b_{1, j_{\text{L}} - 1}		&0								&\cdots	&0				&b_{1, j_0 + 1}					&\cdots	&b_{1, L}	\\ 
\vdots					&		&								&\vdots							&		&				&\vdots							&		&	\\
b_{j_{\text{L}} - 1, 1}	&\cdots	&b_{j_{\text{L}} - 1, j_{\text{L}} - 1}	&0						&\cdots	&0				&b_{j_{\text{L}} - 1, j_0 + 1}	&\cdots	&b_{j_{\text{L}} - 1, L}	\\
0						&\cdots	&0								&b_{j_{\text{L}}, j_{\text{L}}}	&		&0				&0								&\cdots	&0	\\
\vdots					&		&								&								&\ddots	&				&								&		&	\\
0						&\cdots	&0								&0								&		&b_{j_0, j_0}	&0								&\cdots	&0	\\
b_{j_0 + 1, 1}			&\cdots	&b_{j_0 + 1, j_{\text{L}} - 1}	&0								&\cdots	&0				&b_{j_0 + 1, j_0 + 1}			&\cdots	&b_{j_0 + 1, L}	\\
\vdots					&		&								&\vdots							&		&				&\vdots							&		&	\\
b_{L, 1}				&\cdots	&b_{L, j_{\text{L}} - 1}		&0								&\cdots	&0				&b_{L, j_0 + 1}					&\cdots	&b_{L, L}
\end{pmatrix},
\end{align*}
\end{widetext}
where $j_{\text{L}}(t) = j_0 - \bar{v}t$. For $\rho_{\uparrow \text{R}}(t)$, the section corresponding to indices $j_0, \cdots, j_0 + \bar{v}t$ is diagonal. The matrices $\rho_{\uparrow \text{L}}(t)$ and $\rho_{\uparrow \text{R}}(t)$ are now divided into coherent and incoherent (diagonal) blocks as the part of the density matrix corresponding to the sites the impurity has reached becomes diagonal. The function 
\begin{align}
g(t) = \text{Tr}(\rho_{\uparrow}^2(t)), 
\label{eq:gt}
\end{align}
where $\rho_{\uparrow}(t)$ is given by eq. (\ref{eq:density_matrix_up}), is drawn in Fig. \ref{fig:two_particles_function}. Since $\bar{v}t$ is a continuous variable, we have used a discretization of position that is smaller than the lattice spacing in order to make $g(t)$ continuous in the figure.

\begin{figure}[h!]
\begin{center}
	\includegraphics[width=0.49\linewidth]{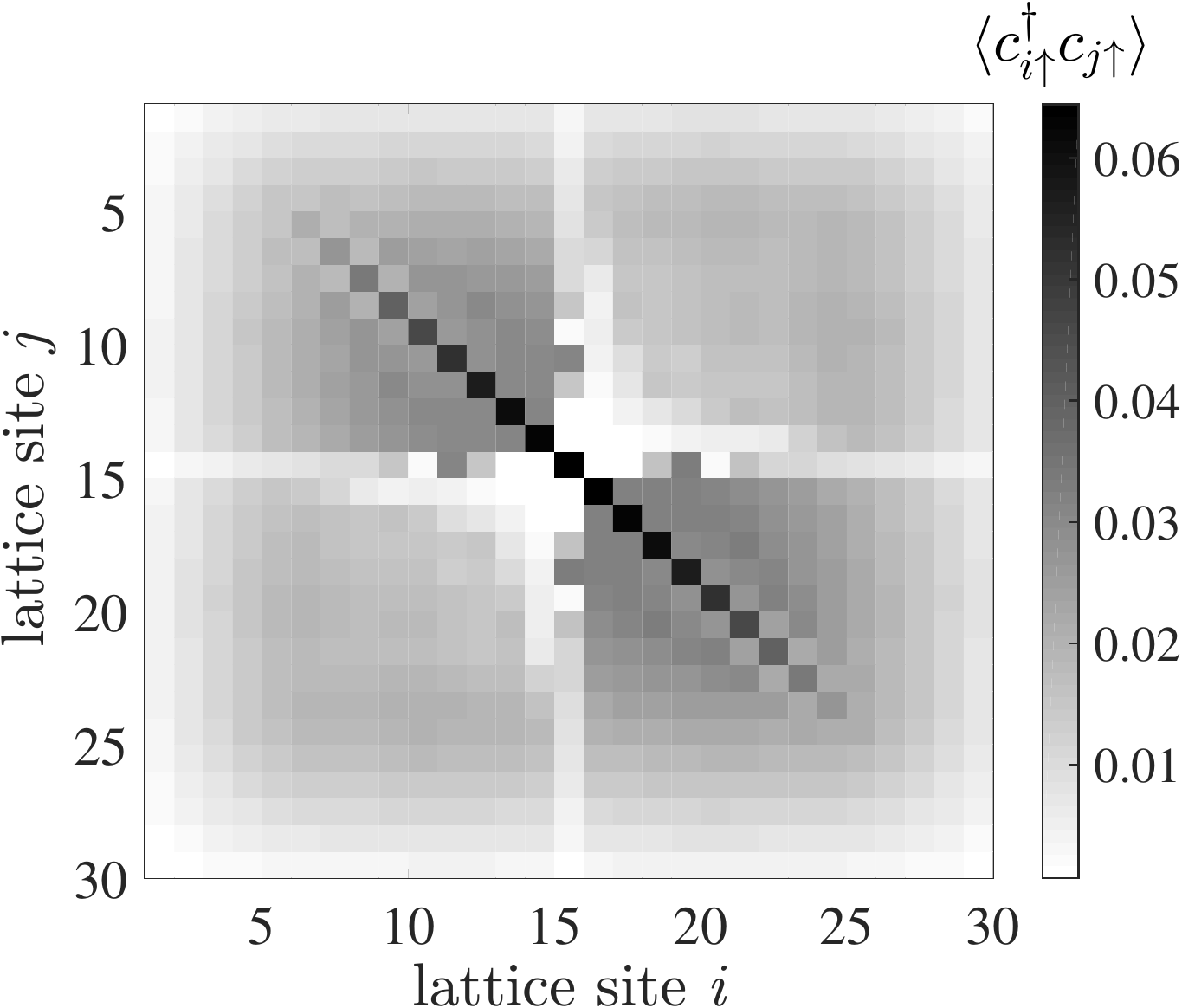}
	\includegraphics[width=0.49\linewidth]{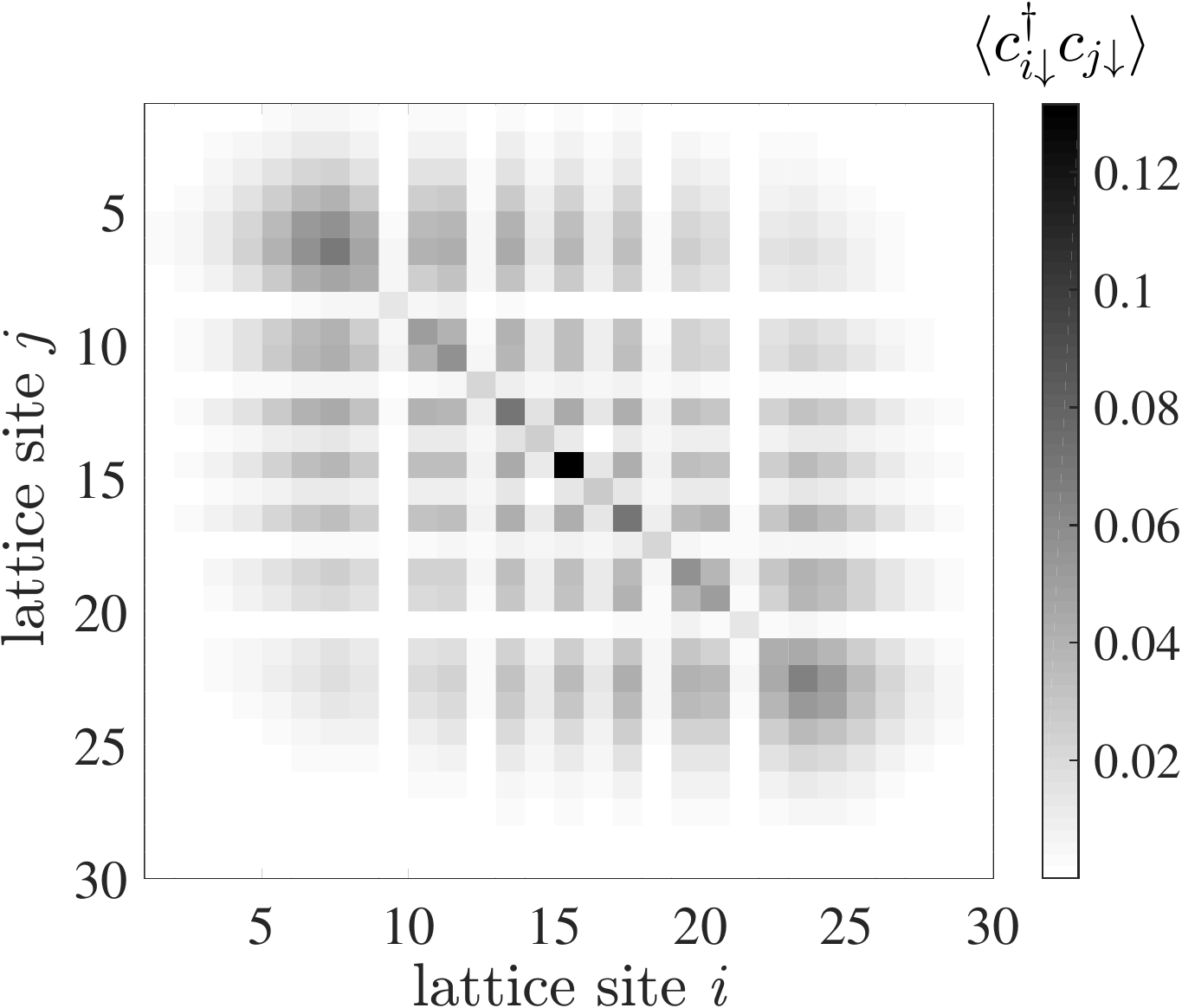}
	\caption{The reduced density matrices of the spin-up (left) and spin-down (right) particles at time $t = 5/J_{\downarrow}$ for $U = 20 J_{\downarrow}$.}
\label{fig:density_matrix_2particles}
\end{center}	
\end{figure}

The left panel of Fig.~\ref{fig:density_matrix_2particles} shows a similar feature of lower off-diagonal values corresponding to the central sites. The reduced density matrix of the spin-down fermion has the largest value at the center, corresponding to the peak in the density distribution of Fig. \ref{fig:line_density_2particles}. A system with weak interactions would not be well described by this model since the particles would be less entangled and there would not be such a nearly diagonal section in the density matrix of the bath. For more bath particles $N_{\uparrow} > 1$, the reduced density matrix of the bath would have larger dimensions and would be more complicated to analyze. We however expect the same underlying effect to be present at low filling fractions. We do not expect this description to hold for strong interactions and large filling since doublon formation is not taken into account here. 

\section{Purities of the reduced density matrices of two subsystems}
\label{app:Schmidt_decomposition}

By the Schmidt decomposition, a pure state $\ket{\psi}$ of a bipartite system with subsystems $A$ and $B$ can be written
\begin{align*}
\ket{\psi} = \sum_i c_i \ket{\phi_i} \ket{\chi_i},
\end{align*}
where $\ket{\phi_i}$ and $\ket{\chi_i}$ are orthonormal bases of $A$ and $B$, respectively. The reduced density matrix of $A$ is
\begin{align*}
\rho_A &= \text{Tr}_B(\rho) = \sum_i \bra{\chi_i} (\ket{\psi} \bra{\psi}) \ket{\chi_i} \\
&= \sum_i \bra{\chi_i} \left(\sum_j c_j \ket{\phi_j}\ket{\chi_j} \sum_l c_l^* \bra{\chi_l} \bra{\phi_l} \right) \ket{\chi_j} \\
&= \sum_i |c_i|^2 \ket{\phi_i} \bra{\phi_i},
\end{align*}
and the reduced density matrix of $B$ is similarly $\rho_B = \sum_i |c_i|^2 \ket{\chi_i} \bra{\chi_i}$. The reduced density matrices are thus diagonal in these bases and the purities are given by
\begin{align*}
\text{Tr}(\rho_A{}^2) = \text{Tr}(\rho_B{}^2) = \sum_i |c_i|^4.
\end{align*}

\section{Numerical error in the total energy}
\label{app:error}

\begin{figure}[h!]
\begin{center}
	\includegraphics[width=0.49\linewidth]{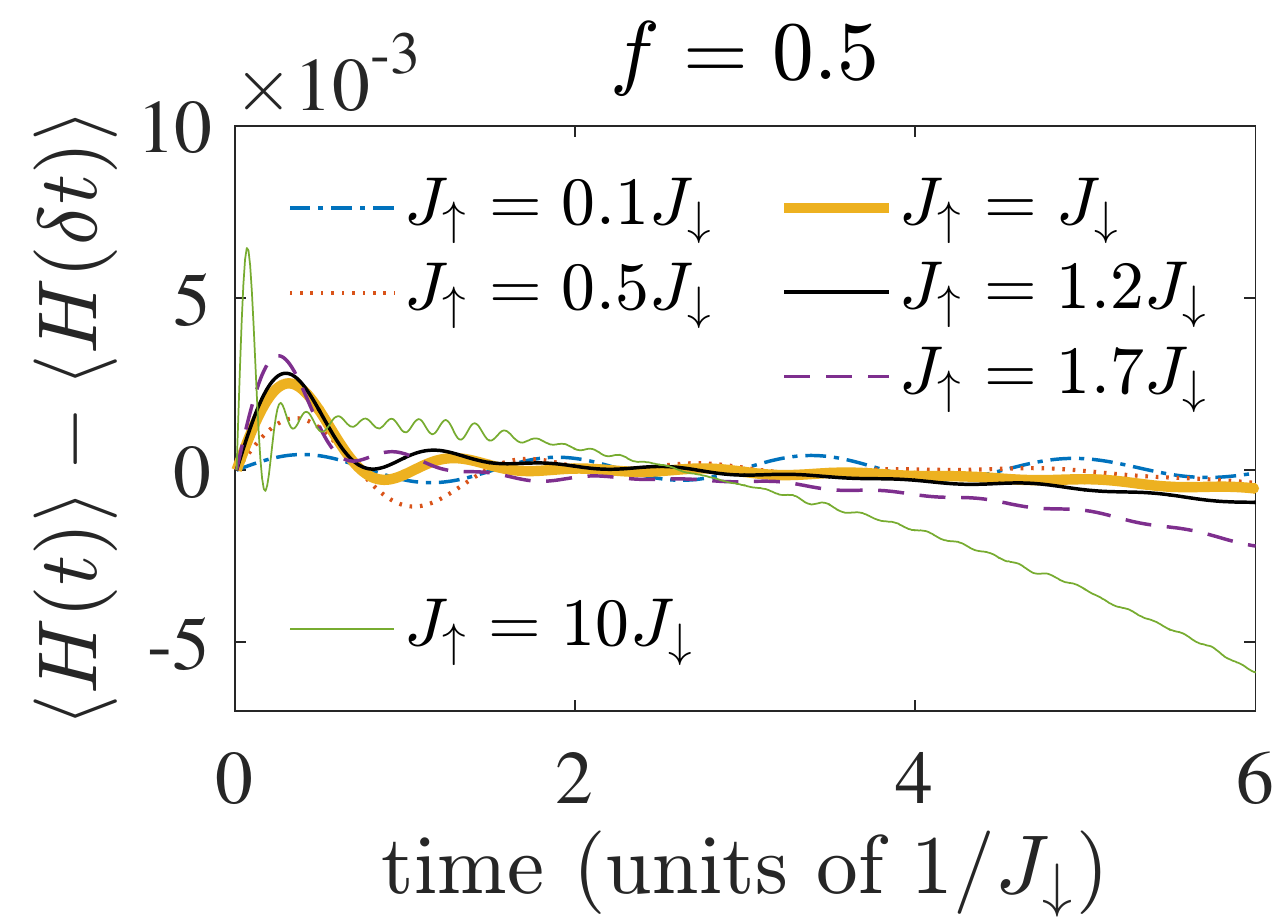}
	\includegraphics[width=0.49\linewidth]{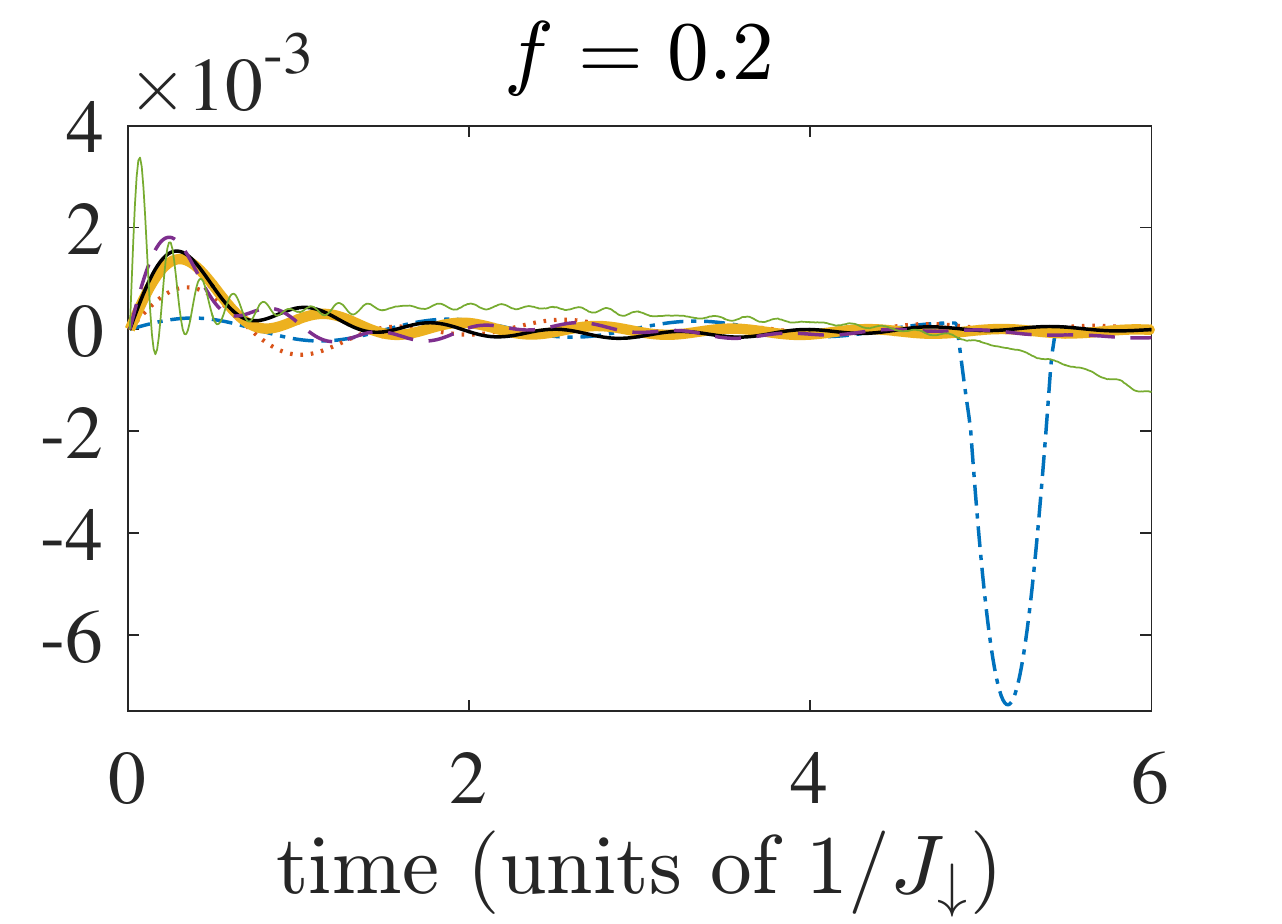}
	\caption{The total energy after the quench $U = 0 \rightarrow 1 J_{\downarrow}$ as a function of time, for $f = 0.5$ (left) and $f = 0.2$ (right). Here, $U = 1 J_{\downarrow}$.}
\label{fig:total_energy}
\end{center}	
\end{figure}

Figure~\ref{fig:total_energy} shows the change in total energy after the first step in the time evolution. As energy is conserved, this quantity should be zero and it now serves as a check for the errors of the numerical method. The small deviations from zero result from the truncation of the state and the discretization of time into finite steps.
 
\bibliographystyle{unsrt}
\addcontentsline{toc}{section}{Bibliography}
\bibliography{bibfile}

\end{document}